\documentclass[12pt,preprint]{aastex}
\usepackage{verbatim}
\usepackage{graphicx}
\usepackage{verbatim}
\usepackage{epstopdf}
\usepackage{rotating}
\usepackage{natbib}
\shorttitle{Retrograde Planets}
\shortauthors{}

\usepackage{color}
\usepackage{ulem}
\newcommand{\ACBc}[1]{\textcolor{black}{ #1}}

\newcommand{\qf}{q_{}}

\begin{document}
\title{Interactions Between Moderate- and Long-Period Giant Planets: Scattering Experiments for Systems in Isolation and with Stellar Flybys}
\author{Aaron C.~Boley, Matthew J.~Payne, and Eric~B.~Ford}
\affil{Department of Astronomy, University of Florida, 211 Bryant Space Science Center, PO Box 112055, Gainesville, FL 32611}

\begin{abstract}
The chance that a planetary system will interact with another member of its host star's nascent cluster would be greatly increased if gas giant planets form  {\it in situ} on wide orbits.    In this paper, we explore the outcomes of planet-planet scattering for a distribution of multiplanet systems that all have one of the planets on an initial orbit of 100 AU.  The scattering experiments are run with and without stellar flybys.  We convolve the outcomes with distributions for protoplanetary disk and stellar cluster sizes to generalize the results where possible.  We find that the frequencies of large mutual inclinations and high eccentricities are sensitive to the number of planets in a system, but not strongly to stellar flybys.  However, flybys do play a role in changing the low and moderate portions of the mutual inclination distributions, and erase dynamically cold initial conditions on average.  Wide-orbit planets can be mixed throughout the planetary system, and in some cases, can potentially become hot Jupiters, which we demonstrate using scattering experiments that include a tidal damping model. If planets  form on wide orbits {\it in situ}, then there will be discernible differences in the proper motion distributions of a sample of wide-orbit planets compared with a pure scattering formation mechanism.  Stellar flybys can enhance the frequency of ejections in planetary systems, but auto-ionization is likely to remain the dominant source of free-floating planets.  
  
\end{abstract}

\section{Introduction\label{sec:intro}}

Observations have revealed a rich distribution of planetary system architectures\footnote{exoplanets.org \citep{wright_etal_2011_pasp_123}}.  Massive Jovian planets can be found on orbits with periods ranging from a few days to thousands of years \citep[e.g.,][]{marcy_etal_1997_apj_481, kalas_etal_2008_science_322,marois_etal_2008_science_322,lafreniere_etal_2010_apj_719}. Super-Earths and Neptune-size planets are abundant within stellar separations of 0.5 AU  \citep{borucki_etal_2011_apj_736}.  Multi-planet systems are common, including densely-packed orbital configurations \citep[e.g., Kepler-11,][]{lissauer_etal_2011_nature_470}. This diversity demonstrates that planets cannot be thought of as isolated objects slowly growing  within their respective feeding zones.  Even in the Solar System, the Late Heavy Bombardment, Kuiper Belt orbital structure, asteroid belt composition, and the mass of Mars suggest that the Solar System planets experienced substantial migration \citep[e.g.,][]{walsh_etal_2011_nature_2011}.

The architectures of large bodies in planetary systems are sculpted by at least two general mechanisms: n-body dynamics and disk-planet interactions.  Neither is mutually exclusive.  Examples of the former include planet-planet scattering \citep[e.g.,][]{rasio_ford_1996_science_274,lin_ida_2007_apj_477}, interactions with a stellar companion \citep[e.g.][]{holman_etal_1997_apj_366}, and planet-planet-stellar perturber excitation \citep{adams_laughlin_2001_apj_150, zakamska_tremaine_2004_aj_128, malmberg_etal_2011_mnras_411}. The resulting scattering could explain the planet eccentricity distribution, for which the median eccentricity $e\approx 0.14$ \citep{wright_etal_2011_pasp_123}, and could even explain highly inclined and in some cases retrograde systems \citep[][]{chatterjee_etal_2008_apj_686,nagasawa_etal_2008_apj_678,triaud_etal_2010_aa_524}. 

\ACBc{The second mechanism, disk-planet interactions, can cause planets to move throughout the nebula \citep{kley_nelson_2012_review}.  Detailed planetary type I migration studies that include proper thermodynamics \citep{paardekooper_mellema_2006_aa_459} and radiation hydrodynamics \citep{kley_etal_2009_aa_506} show that migration can be inward or outward for a range of conditions, with zero-torque radii possible as well.  If two massive planets open a mutual gap in the disk, then their migration can also be inward or outward, depending on the details of a given disk's structure and the planet mass ratios \citep{snellgrove_etal_2001_aa_374,crida_etal_2009_apjl_705}.   Disk-planet interactions typically lead to eccentricity and inclination damping \citep{bitsch_kley_2010_aa_523,bitsch_kley_2011_aa_530} for the majority of planet masses and disk conditions  \citep{moorhead_adams_2008_icar_193}.  Eccentricity excitation may also be possible for large planet masses or specific disk conditions \citep[e.g.,][]{goldreich_tremaine_1980_apj_241,ogilvie_lubow_2003_apj_2003}.  While excited systems may be best explained by n-body interactions, densely-packed systems like Kepler-11 or systems in or near resonances \citep[e.g.,][]{lissauer_etal_2011_apj_inpress} likely require a phase of planet-disk interactions. }

These mechanisms can, either separately or in combination \ACBc{\citep{moorhead_adams_2005_icar_178}}, turn a planetary system with planets on moderate-period orbits, e.g., between $\sim 1$ to 10 AU, into a system with planets on short- and long-period orbits.  Nonetheless, it remains to be seen whether scattering and/or migration can match the constraints set by multi-planet systems \citep[e.g.,][]{veras_etal_2009_apj_696, dodson-robinson_etal_2009_apj_707}.  It is also possible that formation at moderate periods is not the only mode of planet formation, with at least some wide-orbit planets forming {\it in situ} by disk instability during the earliest stages of disk evolution \citep{boss_1997_science_276,boley_2009_apj_695}.  If a massive planet can form at large stellar separations, regardless of the mechanism, then the cross section for significant perturbation of the planetary system by stellar flybys would be much larger than for the solar system, and stellar flybys may be more important in shaping planetary orbits.  For example, a distant stellar flyby could cause otherwise stable systems to grow to instability due to a cascade of eccentricity pumping \citep{zakamska_tremaine_2004_aj_128} or to decrease the decay timescale of the system \citep{malmberg_etal_2011_mnras_411}.  Wide-orbit planets that are placed on highly inclined orbits could also induce Kozai oscillations with other system members \citep[e.g.,][]{naoz_etal_2011_nature_473}.  Finally, just as planets that form at short periods may have scattered onto on wide-orbits, e.g., \citet{veras_etal_2009_apj_696}, planets that form on wide-orbits may be placed on short periods through multiple  scattering events.  

In this paper, we explore outcomes for planet-planet scattering under the assumption that planetary architectures can begin with planets on wide-orbits.  We compare isolated systems with systems that experience stellar flybys.    In Section \ref{sec:encounter_freq}, we discuss encounter likelihoods, and using rates from the literature, estimate the fraction of field stars that have had an encounter with a pericenter less than some value $\qf$.   We then describe our base set of scattering experiments in Section \ref{sec:numerics}.  We present the results in Section \ref{sec:results} and in Section \ref{sec:discussion}, use those results to determine expectation values for median inclination and eccentricities among distributions of field star planetary systems.   We also demonstrate that proper motion distributions can be used to discriminate between formation modes, and show  that planets on initial 100 AU orbits can become hot Jupiter candidates.  We conclude with a summary of the results in  Section \ref{sec:conclusion}. A summary of the symbols used in this manuscript is given in Table \ref{tab:definitions}.  

\section{Encounter Frequency\label{sec:encounter_freq}}

\citet[][]{proszkow_adams_2009_apj_185}, hereafter PA2009, characterized the encounter rates for stars in cluster sizes between $N=100$ and 3000 for a wide range of parameters \citep[see also][]{adams_etal_2006_apj_641}.  In their study, they focused on both virial ($Q=0.5$) and subvirial ($Q=0.04$) velocity dispersions, where $Q=\left|\right.$Total Kinetic Energy/Total Potential Energy$\left|\right.$. For some clusters, they explored the sensitivity of the interaction rate to the star cluster core radius $r_c$ by varying $Q$.  Here, we use the results from their $Q=0.04$ initial conditions (ICs) with a cluster core radius scaling $r_c=1{\rm~pc}~(N/300)^{1/2}$, which we choose for three principal reasons. (1) The velocity dispersion among prestellar cores is observed to be small \citep[e.g.,][]{andre_2002_apss_281}, suggesting that star clusters are out of virial equilibrium at birth. (2) Star cluster cores during their gas-embedded phase are initially compact \citep{bastian_etal_2008_mnras_2008}, and expand to the sizes found by \citet{lada_lada_2003_araa_41} as they evolve, with ambient gas removal likely playing a role in the cluster's expansion \citep{bastian_goodwin_2006_mnras_369}.  PA2009 found that their subvirial ICs give an effective core radius that is $\sim \sqrt{2}$ smaller than the initial $r_c$, which is more inline with the Bastian et al.~results. (3) We are specifically interested in clusters that \ACBc{have short lifetimes and are the dominate contributors to} the field population, which are the targets for most planet discovery surveys.  This limits cluster sizes to be $\lesssim10^4$. Within this parameter space, star cluster core radii follow the Lada \& Lada scaling $r_c\propto N^{1/2}$.

%\begin{table}
%\begin{center}
%\caption{Nomenclature}
%\begin{tabular}{ l l }\\\hline
%$Q$ & Virial parameter \\
%$r_c$ & Star cluster core radius\\
%$N$ & Number of cluster members\\
%$\qflyby$ & Pericenter for perturber\\
%$\Gamma(p,N)$ & Rate of encounters within $p$ in a cluster of size $N$\\
%$\gamma(N)$ & Rate exponent\\
%$\xi(p,N)$ & The number of encounters $<p$ in a cluster of size $N$\\
%$f_m$ & Star cluster mass function\\
%$f_N$ & Star cluster number function\\
%$\eta$ & Fraction of all fields stars that experienced at least one close encounter $<p$\\
%$P$ & Binary period. Used for binary ICs\\
%$f_{i>40^{\circ}}(p)$ & Fraction of systems per pericenter that have one planet with an inclination $i>40^{\circ}$\\
%$\eta_{i>40^{\circ}}$ & Fraction of field stars that have one planet with an inclination $i>40^{\circ}$\\
%$f_{e>0.1}(p)$ & Fraction of systems that have an eccentricity $e>0.1$ per pericenter\\\hline
%\end{tabular}
%\end{center}
%\end{table}

To proceed, we first make a simple estimate as to whether close encounters could be important for producing highly inclined outer planets in the field star population.  Let $\Gamma(\qf,N)$ be the time-averaged rate for all encounters with a stellar flyby pericenter $\le \qf$ in a nascent cluster of size $N$. We roughly model the PA2009 results (their Table 8) using the following:
\begin{equation}
\Gamma(\qf,N) \approx 0.26 \left(\frac{100}{N}\right)^{1/2} \left(\frac{\qf}{1000~{\rm AU}}\right)^{\gamma(N)}{\rm encounters\ per\ star\ per\ Myr}.
\label{eqn:flybyrate}
\end{equation}
 We determined the functional form for $\gamma$ using the tabulated results of PA2009, and the value of $\gamma$ represents the typical degree of gravitational focusing.  As $\gamma\rightarrow1$, gravitational focusing becomes strong, and when $\gamma\rightarrow2$, focusing becomes weak.   We set the rate exponent to  $\gamma(N)=2- \exp\left(-N/782\right)$,  forcing the value of $\gamma$ to be between 1 and 2.   Let $\Delta t$ represent the time period in a cluster during which close encounters remain important, which gives us the number of encounters with a closest approach distance $<\qf$ per star for a given cluster size $N$ as $\Gamma(\qf,N) \Delta t$.  Because $\Gamma$ is averaged  over 10 Myr in PA2009, we will typically take $\Delta t\sim 10$ Myr unless noted otherwise.  
 
 Next, we assume that all field stars come from dissolved clusters with member numbers between $N_0$ and $N_1$.  In this case, we use the canonical star cluster mass function ($m d\xi_m/dm\sim m^{-1}$) to write the star cluster number function; namely,
\begin{equation}
d\xi_N/dN=AN^{-1},
\end{equation}
where $A$ is set to normalize the function to unity.  Finally, we write the average number of encounters per field star for flyby pericenter $<\qf$  as
\begin{equation}
\eta=\int_{N_0}^{N_1} \frac{d\xi_N}{dN}\Gamma\left(\qf,N\right) \Delta t ~ dN \rm .
\label{eq:eta}
\end{equation}
With this definition for $\eta=\eta( \qf, N_0,N_1, \Delta t, \frac{d\xi_N}{dN})$, extra weight will be given to stars that have multiple encounters for pericenters $<\qf$.  We account for this weighting by introducing $\eta'$, which has the same form as $\eta$, but forces $\Gamma \Delta t \le 1$.  The value of $\eta'$ thus represents the fraction of field stars that have had at least one encounter.  The average number of encounters among field stars that have had at least one encounter is given by the ratio of $\eta$ to $\eta'$.
In Table \ref{table:encounter_probabilities}, we give the results of integrating equation (\ref{eq:eta})  over several values of $N_0$ and $N_1$ for $\qf=100$, 200, 300 AU, and 1000 AU, with $\Delta t=10$ Myr.  The results are fairly sensitive to $N_0$, owing to the increased likelihood of a star to have a close encounter in small $N$ clusters, but we do find that 20-40\% of field stars should have experienced at least one encounter within 300 AU.  Next, we discuss the effects of the core cluster size.

\subsection{Sensitivity of Results to Assumptions for Nascent Cluster Core Sizes\label{sec:sensitivities}}

The dominant source of uncertainty in the results for the following calculations is the nascent cluster stellar density during which most collisions occur.   To understand this sensitivity,  we consider a general flyby rate $\Gamma=nv\sigma$, where $\sigma$ is the cross section for the \ACBc{for a star to pass within a distance $q$ of another star}, $n$ is the typical stellar density in the cluster, and $v$ is the typical speed of a star in the cluster.  Let us approximate $n\approx 3N/(4\pi r_c^3)$, where $r_c$ is the cluster core radius, and $v^2\approx GNm/r_c$, where $m$ is the characteristic stellar mass.  We also assume that the \citet{lada_lada_2003_araa_41} cluster size relation holds, simply scaled to higher densities, where $r_c\approx r_0 (N/300)^{1/2}$.  Finally, gravitational focusing must be included in the definition of the cross section, such that $\sigma=\pi q^2 \left[1+0.23 r_0/(q N^{1/2}) \right]$, where $q$ is the largest pericenter considered.  Combining these relations, we find 
\begin{eqnarray}
\Gamma & \approx & 3\times10^{-3} \left(\frac{ 1{\rm~pc}}{r_0}\right)^{7/2}\left(\frac{300}{N}\right)^{1/4}\left(\frac{q}{{1000\rm~AU}}\right)^2\\\nonumber
& &\times  \left(1+3 \left(\frac{r_0}{1{\rm~pc}}\right)\left(\frac{1000{\rm~AU}}{q}\right)\left(\frac{300}{N}\right)^{1/2} \right)\rm~per~star~per~Myr. 
\end{eqnarray}
The above rate is consistent to within a factor of three of the PA2009 rate for their virial $N=300$ cluster.  It shows the limiting behavior of $\gamma(N)$ and that the overall dependence of $\Gamma$ on $r_0$ is quite strong. As discussed in Section 2, we use the encounter rates from the PA2009 ICs that begin out of virial equilibrium, which reduces the effective $r_0$ from 1 to about 0.7 pc.  Using the above arguments, we would expect that an initial cluster core scaling of  0.7 pc would have encounter rates that are about 3.5 times the $r_0=1$ pc rates for weak gravitational focusing.  In comparison,  PA2009 found that their non-virial (cold) simulations were enhanced by a factor $\sim8$ over their virial conditions, which is larger than we would expect.  While starting with dynamically cold ICs leads to a smaller effective cluster size, it is not strictly the same as starting with a more compact cluster.  For example, PA2009 attribute the additional enhancement  to a larger fraction of bound cluster members in the runs with non-virial ICs, ultimately boosting the encounter rate.    Altogether, their cold ICs give a similar boost to the encounter rate for a cluster size of 1 pc that one would expect for $r_0\sim0.55$ pc, assuming weak focusing.  

\section{Numerical Experiments\label{sec:numerics}}

The effect that a close encounter will have on a planetary system depends on the pericenter of the encounter \citep[e.g.,][]{adams_laughlin_2001_apj_150,adams_etal_2006_apj_641}. The odds of making significant changes to a planet with an orbit of, say, 1 AU due to a stellar encounter alone are very small, as a small pericenter is required for the planet to be strongly perturbed.  However, as shown in Table \ref{table:encounter_probabilities}, planets on wide orbits, i.e., with semi-major axes $a\sim 100$ AU, stand a reasonable chance of being strongly perturbed.   A close encounter with a system that has a planet or substellar companion on a wide orbit could cause a scattering cascade in a multiple-planet system, in the same spirit as investigated by \citet{zakamska_tremaine_2004_aj_128}, or produce an inclined outer gas giant/brown dwarf that could then cause Kozai oscillations on an inner planet.  For these reasons, we have designed seven sets of simulations to explore the consequences of close encounters on the inclinations of planets on wide orbits and how these planets interact with other system members.

\subsection{Scattering Experiment Design\label{sec:scattering_design}}

\subsubsection{N-Body Method}\label{NB:Method}
We use the Bulirsch-Stoer \citep[e.g.,][]{numerical_recipes} integrator in the Mercury package \citep{chambers_1999_mnras_304}  to evolve realizations of five different system ensembles.  Simulations are evolved for $10^8$ yr  \ACBc{ and the typical energy error is $\sim 10^{-8}$(see appendix)}. The results obtained using these Bulirsch-Stoer integrations were also independently verified using a hybrid integrator from the same Mercury package, as well as using the GPU-based SWARM\footnote{www.astro.ufl.edu/$\sim$eford/code/swarm} integrator using an Hermite integration scheme. The results obtained using all methods were qualitatively similar, but the Bulirsch-Stoer results were ultimately preferred due to their overall ability to conserve energy during the multiple close planetary-scattering events over the course of the $10^8$ year integration.

\subsubsection{N-Body Initial Conditions}\label{NB:ICs}

All systems have a 1 M$_{\odot}$ primary star and a wide-orbit planet with an initial semi-major axis of 100 AU.  Two ensembles have three planets distributed inside the wide-orbit planet, and two ensembles have two planets inside the wide-orbit planet.  For each of these cases, one ensemble has an incoming perturbing star. We set the perturber to have a stellar mass of 0.3 $M_{\odot}$, set its initial velocity to 1 km/s, and place it randomly on the sky, as seen from the given planetary system, at a distance of 0.1pc. The perturber reaches its pericenter after $\sim 10^5$ yr of evolution. To distinguish between the four cases, we adopt the following nomenclature: 3P1F1 refers to three planets interior to one wide-orbit planet with a perturber (flyby). 3P1F0 refers to a similar system, but with no flyby.  2P1F1 and 2P1F0 follow the same pattern.  For reference, these names, as well as three additional simulations to be described later in the manuscript (3P1F1C, 2P1F1TD, and 3P1F1TD), can be found in Table \ref{tab:definitions}. 

We perform 1,000 realizations of each of the 2P1F0 and 3P1F0 cases, and 3,000 realizations of each of the 2P1F1 and 3P1F1 cases.  In the flyby simulations, the larger sample sizes ensure that we can accurately probe both large- and small-pericenter flybys. 

\begin{figure}
\begin{center}
\includegraphics[width=5cm]{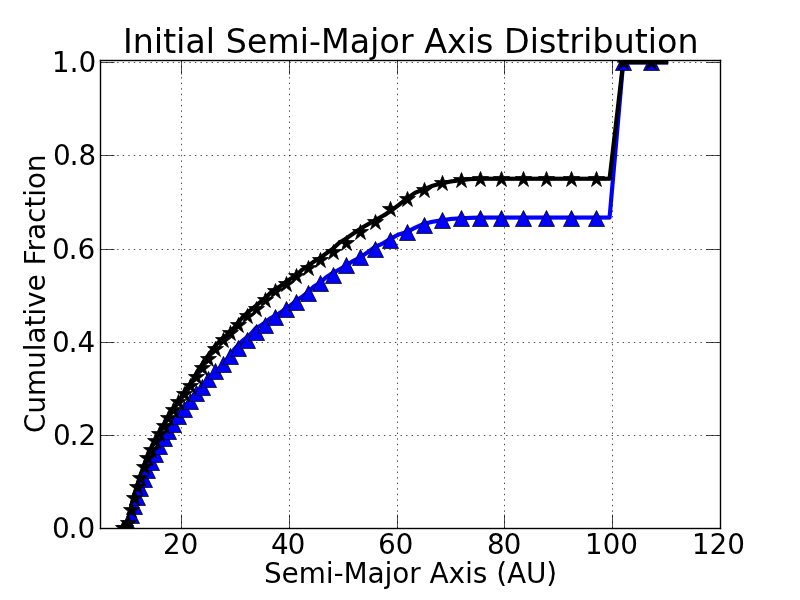}\includegraphics[width=5cm]{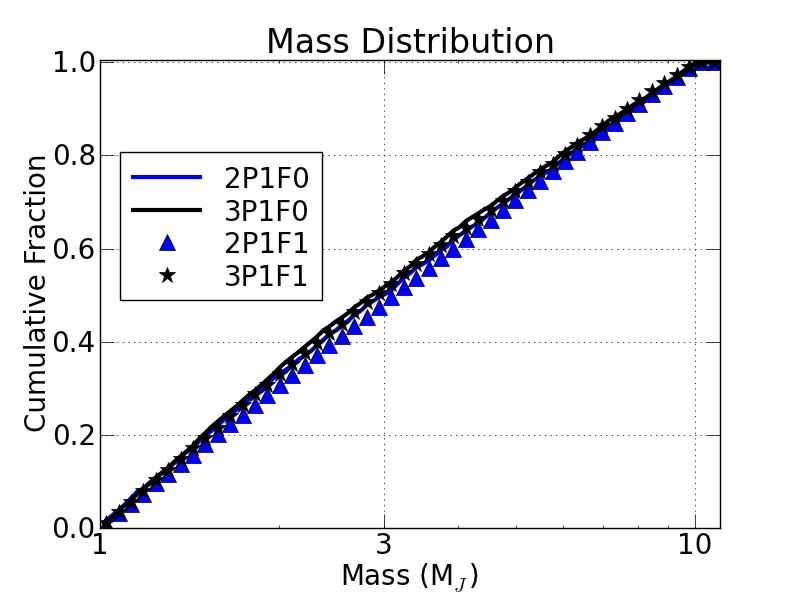}\includegraphics[width=5cm]{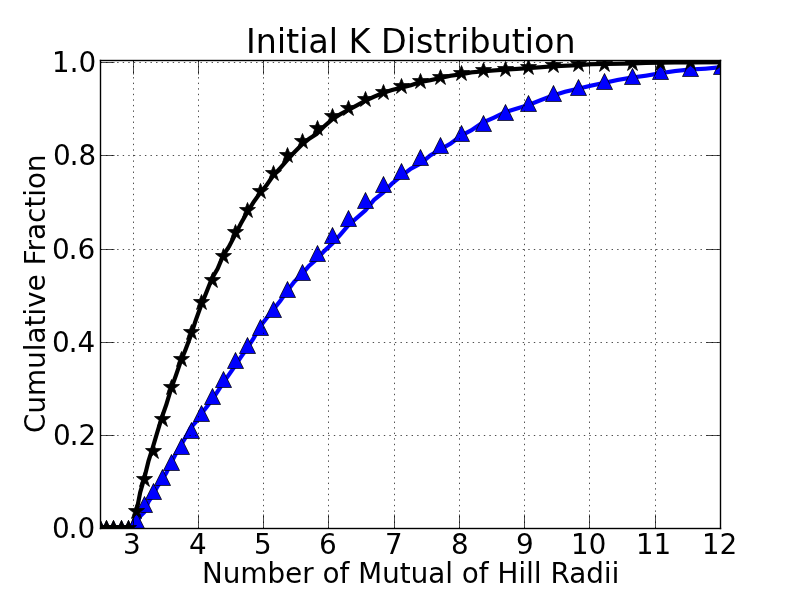}
\caption{Cumulative distributions of the initial conditions for semi-major axis (left), mass (center), and planet orbital separation in number of mutual Hill radii $K$ (right).}
\label{fig:ic_cums}
\end{center}
\end{figure}

All planets have masses drawn uniformly in log space between 1 and 10 $M_J$.  \ACBc{While disk instability may produce a mass distribution that is more top heavy than assumed here, we are not strictly requiring that the formation mechanism must be disk instability.  Moreover, the outcome of disk fragmentation is  an active area of research, and the distribution of fragments that survive to become gas giants, brown dwarfs, or even stars, is not yet known \citep{boley_2010_icar_207,kratter_etal_2010_ApJ_710,nayakshin_2010_MNRAS_408L,zhu_etal_2012_ApJ_746}. }  Planets interior to the wide-orbit planet are given a random inclination, uniformly distributed between zero and $0.1^{\circ}$, and all eccentricities are less than $10^{-3}$.  All wide-orbit planets have zero initial inclination with respect to the $x-y$ plane.  This plane is also taken to be normal to the stellar spin. Planet positions interior to 100 AU are placed randomly, uniform in $a$ and in phase, but with the constraint that any new planet  must be more than three mutual Hill radii from any neighboring planet and must have a semi-major axis $a>10$ AU.  We take the mutual Hill radius $R_H = 0.5 (a_1+a_2)\left[ \left(m_1+m_2\right)/3\right]^{1/3}$ for semi-major axes $a$ and masses $m$, in stellar mass units, for planets 1 and 2.   Figure \ref{fig:ic_cums} displays cumulative distributions for the initial semi-major axes, planet masses, and $K$, the number of mutual Hill radii between any two planets. Three-planet systems that have initial planet-planet spacings $K<3$ exhibit strong interactions on timescales comparable to $\sim10$ orbits of the innermost planet orbit.  The long-term stability of a system rises sharply for $K>3$ \citep[see Appendix B of][]{chatterjee_etal_2008_apj_686}.  Because all systems have the same inner and outer bounds, the 2P1F0 and 2P1F1  systems are less tightly packed than the 3P1F0 and 3P1F1 systems.  

The target distribution of pericenters for the perturber is set to be flat between 0 and 1500 AU.  A flat distribution is biased, overall, toward more frequent close encounters than given by the results of  PA2009. We will account for this bias in Section \ref{sec:discussion}. This sampling is intended to provide better statistics for rare events.  We do note that a roughly flat distribution is expected for small clusters (see $\gamma$  functional form), so this biasing is largest for the largest of clusters.  The distributions were extended to 1500 AU to ensure that we capture weak effects of flybys on systems.  To verify that our calculations properly account for the bias in the $q$ distributions, we also run one set of simulations with a $q$ distribution given by $\gamma=1.3$ between 0 and 1000 AU, which we call 3P1F1C.   Recall that the shape and magnitude of the flyby frequency is dependent on the cluster member number $N$.  For $\gamma\approx 1.3$, the distribution corresponds to an $N= 300$ cluster, with $\eta=1$ at 1000 AU after about 8 Myr.  

\begin{figure}
\begin{center}
\includegraphics[width=8.5cm]{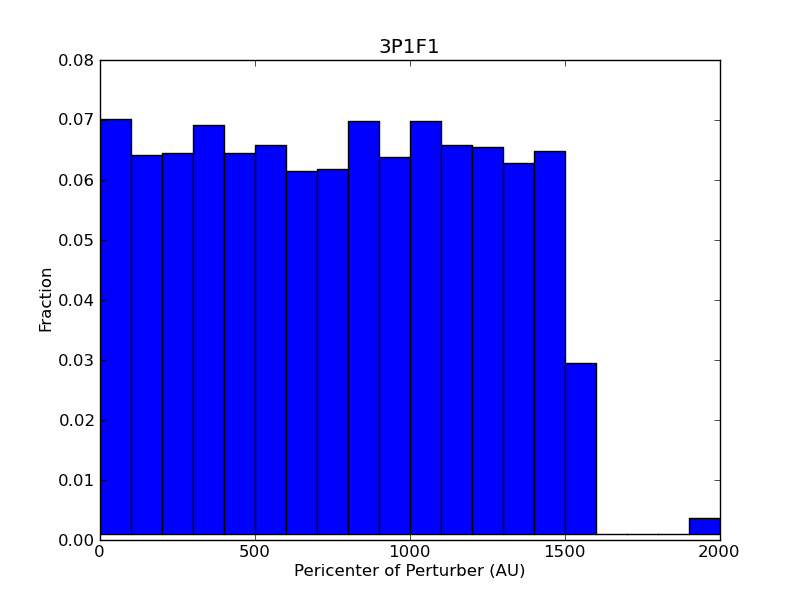}\includegraphics[width=8.5cm]{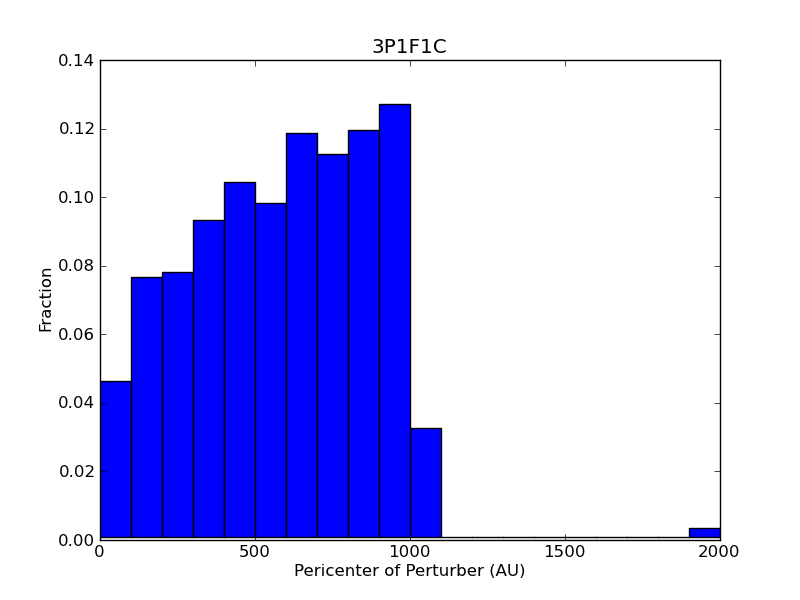}
\caption{Histogram for the pericenters of the perturber $q$.  A few systems extend beyond the cutoffs of 1500 and 1000 AU for the flat ($\gamma=1$) and cluster ($\gamma=1.3$) distribution, respectively.  These systems are included in the rightmost bin shown here. The histogram for 2P1F1 (not shown) is very similar to 3P1F1.  }
\label{fig:hist_peri_flyby}
\end{center}
\end{figure}

The actual pericenter distributions for the perturbers in simulations in 3P1F1 and 3P1F1C are given in Figure \ref{fig:hist_peri_flyby}.  The distribution for 2P1F1 is very similar to 3P1F1, so it is not shown.  There are a few systems in which the flyby has a large pericenter, but as we will show, their evolution will not be altered significantly compared with flybys at 1500 AU.   \citet{adams_laughlin_2001_apj_150} found that the Solar System gas and ice giants will have their eccentricity or mutual inclinations doubled for \ACBc{flybys of binaries within a cross section of (400 AU)$^2$, where the perturber masses and binary orbits were drawn from measured distributions (see paper for details).   They also found that the  cross section for ejections due to flybys is $\sim (73\ {\rm AU})^2$. }   Scaling the cross section of the Solar System to the size of systems studied here, we expect to sample  a full range from weak to very strong interactions.

\section{Results\label{sec:results}}

In this section, we present results from the scattering experiments.  We investigate the inclination distribution for each ensemble, and then explore the degree of \ACBc{radial} mixing of planetary orbits\ACBc{, i.e., large changes in planets' semi-major axes.}   We also explore rare but non-negligible results, such as making hot Jupiters from planets that started at $a\sim 100 $ AU, as well as the effects of flybys on the dynamical stability of these systems. We compare the simulations with and without flybys in several ways. 
First, we investigate differences between the simulations with and without flybys, keeping in mind the following caveats: (1) interactions with low $\qf$ 
will in general be overemphasized in the raw distributions. This is partially offset, however, by (2) the extension of the flyby distribution well beyond $\eta=1$ for $\Delta t=10$ Myr.  For example, an $N=300$ cluster with $\gamma\approx 1.3$, has an $\eta=1$ surface at about $\qf\approx 750$ AU after $\Delta t= 10 $ Myr.  Giving equal weight to flybys out to 1500 AU will  tend to deemphasize the effects of flybys.  Second, we show raw distributions with the data clipped to include only $\qf > 300 $ AU because we expect flybys to have their strongest effects on a system when $\qf\lesssim2.5 R_{\rm sys}$, where $R_{\rm sys}$ is the orbital distance of the outer planet in a given system (see section \ref{sec:extreme_outcomes}).  This selects moderate to weak interactions.  
Third, we weight each system's contribution to the cumulative distribution by the corresponding $\qf$, using $\qf^{\gamma-1}$ from equation (\ref{eqn:flybyrate}), to address results for realistic flyby frequencies. We assume $N=300$ and $\Delta t=10$ Myr, and select only flybys with $\qf<750$ AU.  Finally, we include the results of 3P1F1C, which experienced a realistic flyby distribution for clusters of $N\sim300$.

\subsection{Inclination Variations}
\begin{figure}
\begin{center}
\includegraphics[width=8.5cm]{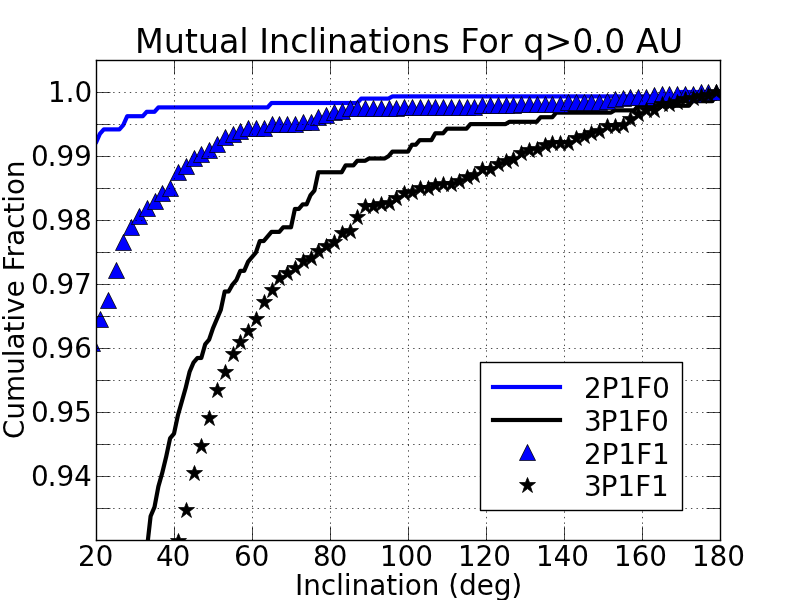}\includegraphics[width=8.5cm]{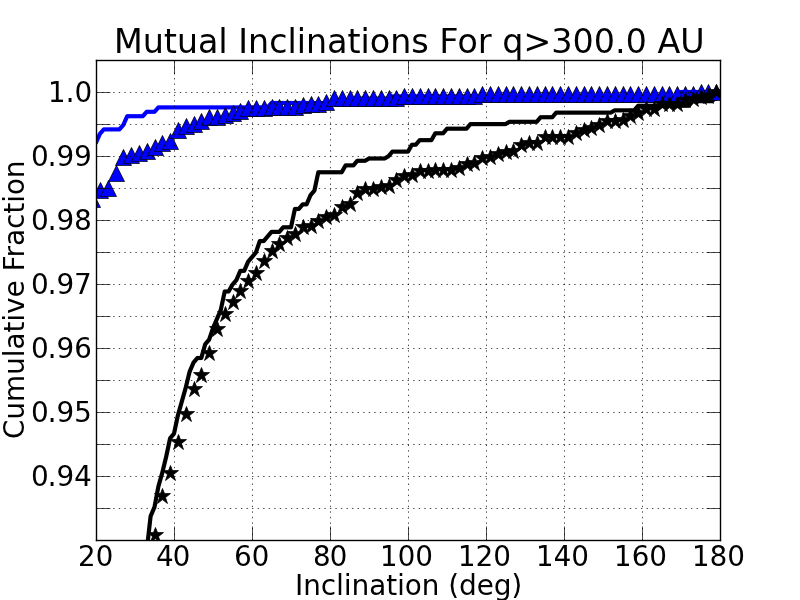}
\caption{Raw cumulative distributions for the end-state mutual inclinations of all planets for all systems that had a $\qf>0$ and 300 AU in the left and right panels, respectively.  Simulations 2P1F0 and 3P1F0 are the same in each plot because no stellar flyby occurred. The maximum mutual inclination is taken for each planet.}
\label{fig:hist_i_mutual_qcut}
\end{center}
\end{figure}

Raw cumulative distributions for the mutual inclinations of all planets at the end of the simulations  ($10^8$ yr) are shown in Figure \ref{fig:hist_i_mutual_qcut}, where we show the maximum mutual inclination between each pair of planets in each system.  Note that these plots do not show the inclination relative to some fixed plane as might be considered in measurements of the Rossiter-McLaughlin (RM) effect.  We will address RM measurements in Section \ref{sec:extreme_outcomes}. 
 The distributions for simulations 2P1F0 and 3P1F0 are the same in each panel because no stellar flyby occurred.    
 Perturbations from passing stars increase the fraction of planets on mutual inclinations $i>40^{\circ}$ \ACBc{by an additional 2\% of all systems for 3P1F1 and an additional one percent of all systems} for 2P1F1. While these changes are relatively small compared with the entire distribution, the fraction of systems with large inclination planets is doubled to tripled in 2P1F1 compared with 2P1F0 and is increased by 40\% in 3P1F1 compared with 3P1F0.  When only encounters with pericenters $\qf>300$ AU are considered, the fraction of planets with $i>40^{\circ}$ is indistinguishable between simulations.    For retrograde orbits, there is the possibility of an enhancement by flybys for the 4-planet systems, as the fraction of systems doubles from $\sim 1\% $ to $\sim 2\%$.  Caution must be taken, though, as the results rely on variations that are similar to the expected noise between the samples.  Figure \ref{fig:hist_i_mutual_qcut} also demonstrates that 
an extra planet in the system (2P1F0 compared with 3P1F0) raises the high-inclination distribution by a factor of 10 (from $\sim 0.5\%$ to $\sim 5.5\%$) at $40^{\circ}$.   While this is a relatively small increase compared with the entire population, the fraction of planets that could effect Kozai oscillations is increased significantly. 
An extra planet in the system is much more important for producing planets with high mutual inclinations than are flybys, even if the system has a planet on an initial semi-major axis of 100 AU. 

\begin{figure}
\begin{center}
\includegraphics[width=8.5cm]{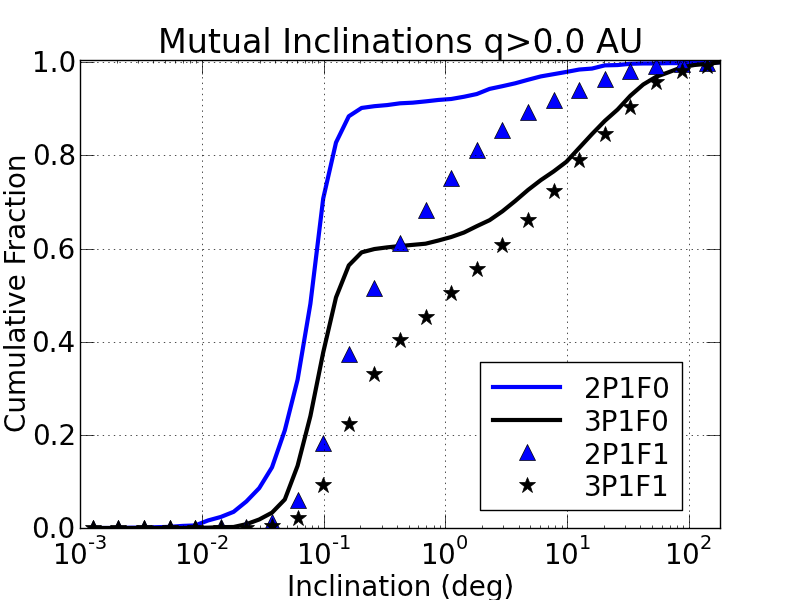}\includegraphics[width=8.5cm]{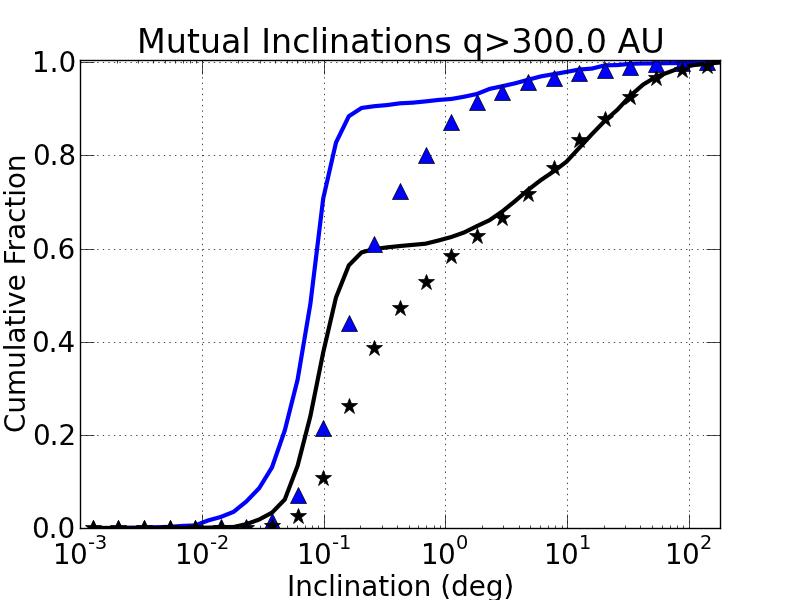}
\caption{  Similar to Figure \ref{fig:hist_i_mutual_qcut}, but with logarithmic inclinations bins to allow all inclinations to be compared.  Simulations without flybys have two components in the distribution, with one reflecting the initial conditions at the other a high-inclination, scattered component.  Flybys blend the peaks in the profile and shift the median inclination to higher values, even for $\qf>300$ AU.  See Table \ref{table:median_inclinations} for median inclinations.     }
\label{fig:distribution_i_mutual}
\end{center}
\end{figure}

In Figure \ref{fig:distribution_i_mutual} we replot Figure \ref{fig:hist_i_mutual_qcut} using a logarithmic inclination axis, showing the full cumulative distributions for the mutual inclinations at the end of the simulations.  
As in Figure \ref{fig:hist_i_mutual_qcut}, two different cuts for $\qf$ are shown.  In the ensembles without a perturber, the distribution has two clear components, with one reflecting the initial conditions of these models and another representing a broad, scattered population.  The medians for the entire distributions are 0.073 and 0.085 degrees for 2P1F0 and 3P1F0, respectively.  The median values for the inclination distributions  with respect to the $x-y$ plane (not shown) are about a factor of two lower, with 0.024 and 0.034 degrees for 2P1F0 and 3P1F0, respectively.  
 %Recall from section \ref{sec:scattering_design} that all simulations are initialized with inclinations distributed between 0 and 0.1 degrees, with respect to the $x-y$ plane, except for the planets at 100 AU, which all have zero initial inclination.  This places the median inclination with respect to the $x-y$ plane for the ICs to be less than 0.5$^{\circ}.  
 For the subset of the distributions $i>0.3^{\circ}$, which selects systems that have had strong planet-planet interactions, the median mutual inclinations  are 3.5 and 11 degrees for 2P1F0 and 3P1F0, respectively, while they are 2.4 (2P1F0) and 5.9 (3P1F0) degrees with respect to the $x-y$ plane.   Table \ref{table:median_inclinations} summarizes these results and lists the initial median inclinations for comparison. In contrast, the distributions for simulations with flybys have a broad distribution from low to high inclinations.   Even selecting only $\qf>300$ AU does not erase this difference. The median mutual inclinations are 0.19 and 0.45 deg for 2P1F1 and 3P1F1, respectively.

\begin{figure}
\begin{center}
\includegraphics[width=8.5cm]{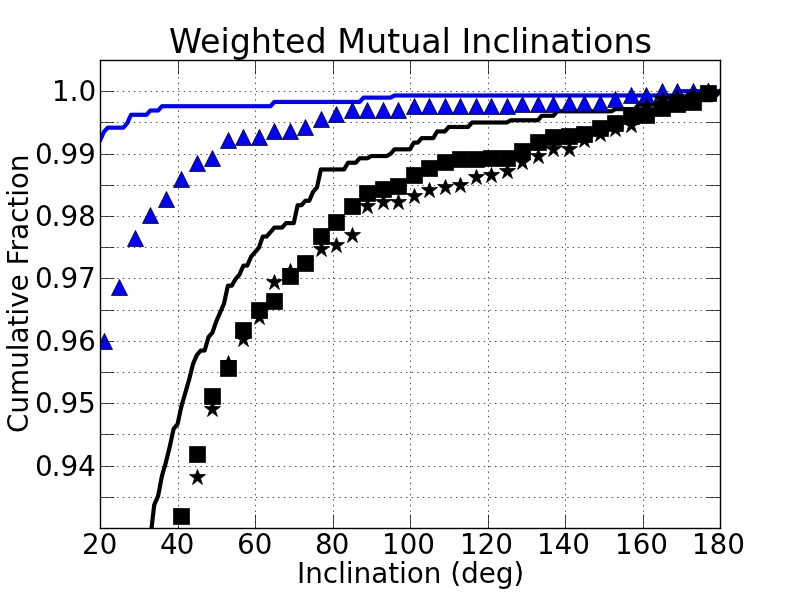}\includegraphics[width=8.5cm]{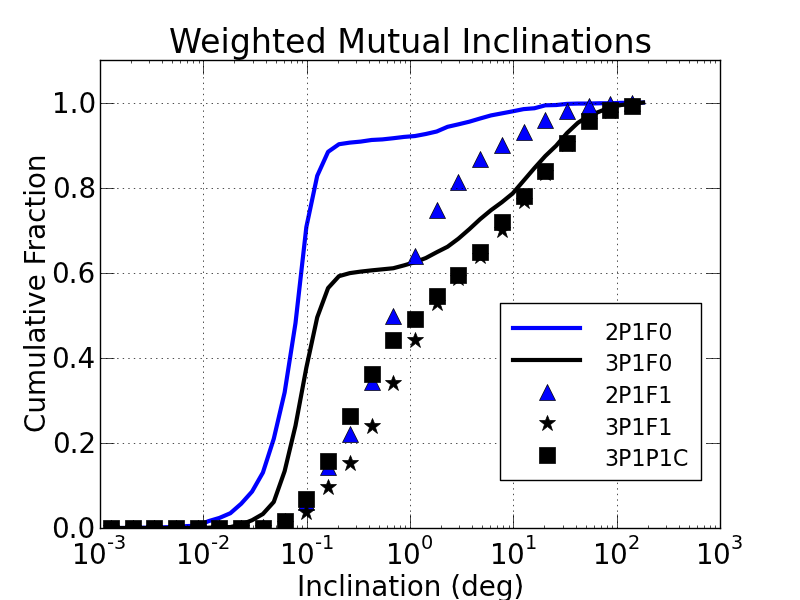}
\caption{Cumulative distributions for the mutual inclinations of all planets for all systems at the end of the simulations, weighted to account for the expected distribution of $\qf$. The initial number of planets or planet orbital density is the primary determinant for the number of highly inclined planets.  Nonetheless, flybys still have an effect on low inclinations.  Even if planets are born perfectly coplanar, the birth cluster of the system will result in an intrinsic inclination spread.  To weight 2P1F1 and 3P1F1, each system's contribution to the histogram is scaled by $\qf^{\gamma-1}$. The results for 3P1F1C are also shown (no weighting necessary), and are consistent with the unweighted 3P1F1 distribution in Figure \ref{fig:hist_i_mutual_qcut}. }
\label{fig:hist_i_weight}
\end{center}
\end{figure}

The distributions in Figures \ref{fig:hist_i_mutual_qcut} and \ref{fig:distribution_i_mutual}  can be weighted to reflect the inclination distributions for the expected perturber pericenter distributions of an $N$-member cluster.  As discussed above, this weighting is dependent on the assumed cluster size, which we take to be $N=300$.   The results  are shown in Figure \ref{fig:hist_i_weight}. The distributions reflect the end-state inclinations.  As seen for the $\qf>0$ AU cuts in Figures  \ref{fig:hist_i_mutual_qcut} and \ref{fig:distribution_i_mutual}, flybys have a noticeable but small  effect on the frequency of planets with mutual inclinations $i>40^{\circ}$ when compared with the entire distribution.  For smaller inclinations, in contrast, the cluster environment has a much stronger effect, and will tend to erase very cold initial conditions. Strict coplanarity should not be expected even in the absence of planet-planet scattering. 
 The results for 3P1F1C are plotted along with the weighted distributions.  The actual cluster distribution is very similar to the weighted one, both of which are similar to the full, flat distribution.  
 
\begin{figure}
\begin{center}
\includegraphics[width=8.5cm]{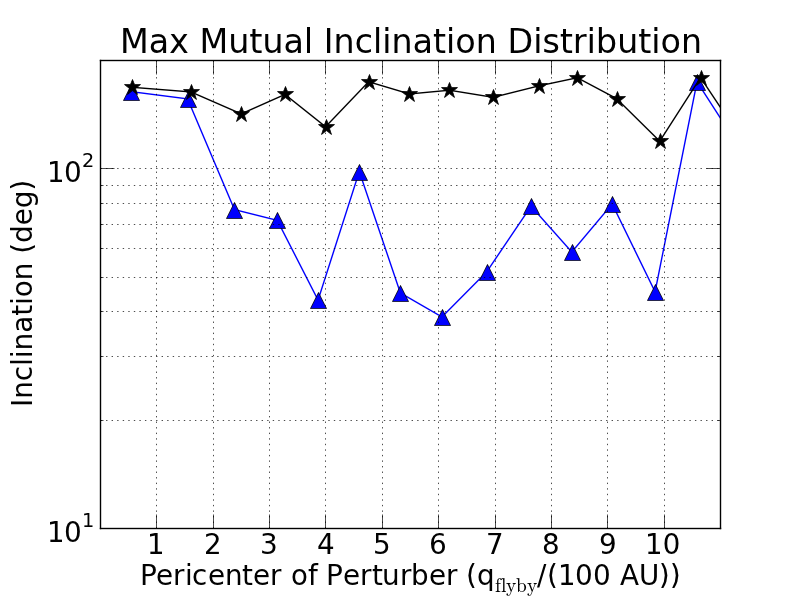}\includegraphics[width=8.5cm]{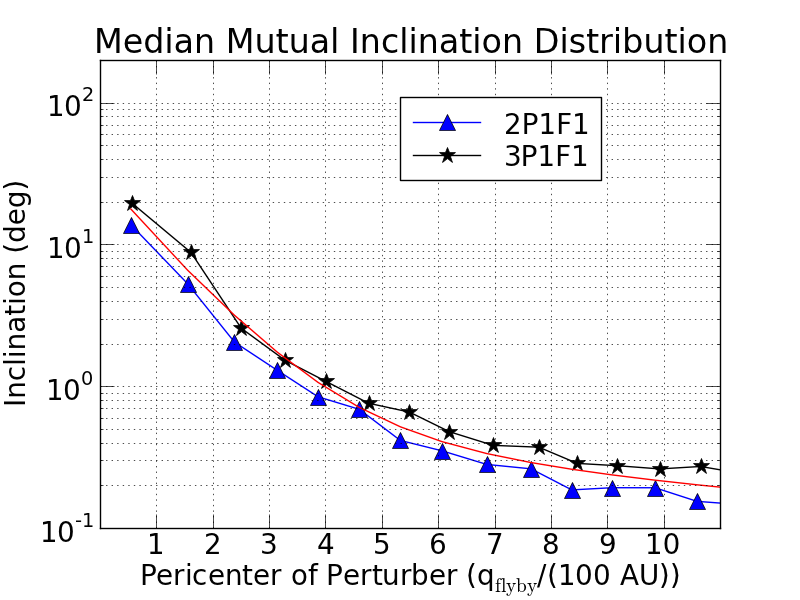}
\includegraphics[width=8.5cm]{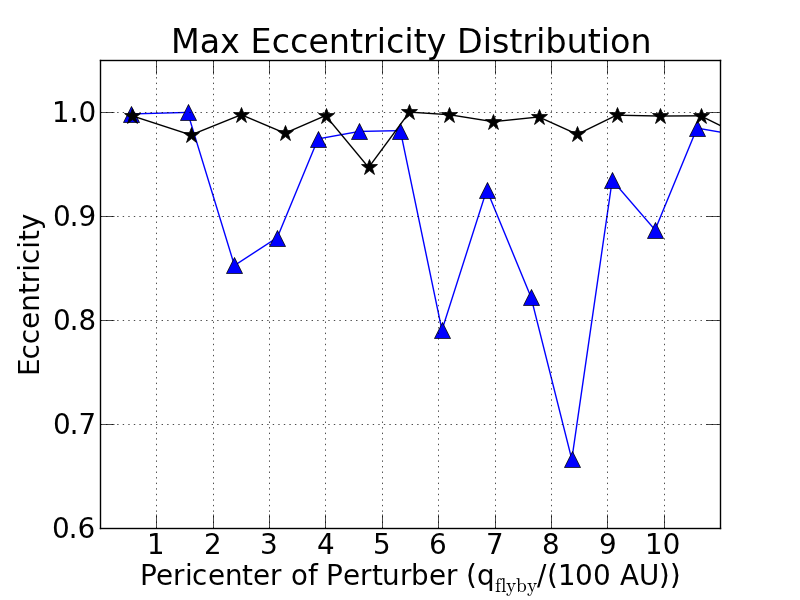}\includegraphics[width=8.5cm]{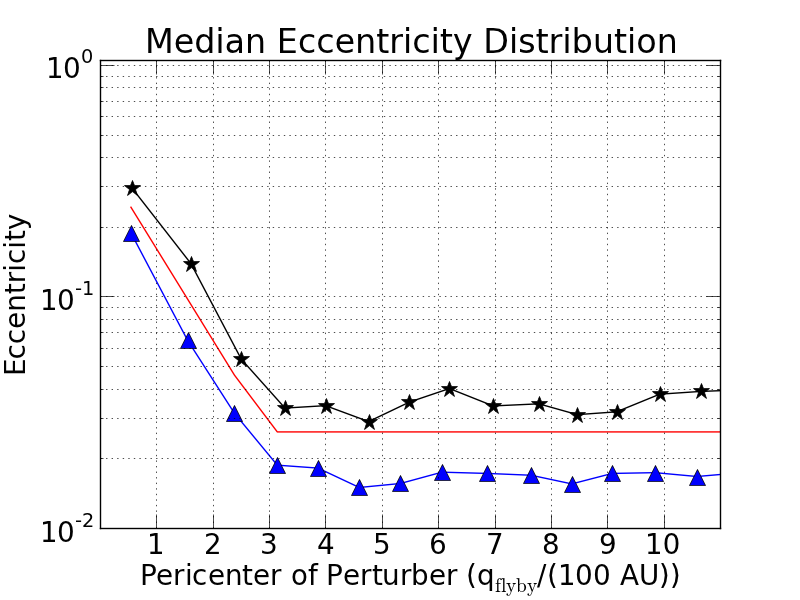}
\caption{Top: The absolute maximum mutual inclination of all systems per $\qf$ bin (left) and the median mutual inclination of all systems per bin.  Flybys can alter inclinations for $\qf$ at least out to 10 times the radial extent of a system. Bins are determined by demanding an equal number of systems per bin.  The most distant $\qf$ that do not form a full bin are not shown.   Bottom: similar as in the top row, but for the maximum and median eccentricity distributions. The maximum values can reflect entirely internal processes, i.e., planet-planet excitation and scattering, while the median values do not.  Fits to the data are shown by the red curves and are described in Section \ref{sec:excitation_discussion_100} }

\label{fig:median_by_pericenter}
\end{center}
\end{figure}

In Figure \ref{fig:median_by_pericenter}, the medians (right) of the eccentricity (bottom) and of the mutual inclination (top) as a function of $\qf$ are shown.  For each planet, the maximum mutual inclination is used, as done in the previous distributions.  In addition, we show the maximum (left) eccentricity and mutual inclination  for all systems in a given $\qf$ bin.  Bin widths are determined by holding the number of systems per bin constant. The most distant $\qf$ that do not form a full bin are not included.  The symbols on the curves correspond to the median $\qf$ for each bin and are placed along the abscissa at the center of the bin width.  The maximum inclination is sensitive to the number of planets/planet orbital density, and shows no dependence on $\qf$ for 2P1F1 or 3P1F1.  The median mutual inclination, in contrast, is not strongly dependent on the density of planets, for the cases studied here, but is dependent on $\qf$.  The 2P1F1 and 3P1F1 distributions for the median mutual inclinations both follow roughly
$25\exp(-x )+2.15/(0.1+x)$ degrees, where $x=q/(100{\rm\ AU})$, over the range shown (shown in red). The median  inclination rises above one degree for flybys that are within 4 times the radial extent of the planetary system, and all median inclinations for all $\qf$ are larger than the medians in the simulations without flybys (see Table \ref{table:median_inclinations}).

 As seen in the inclinations, the maximum eccentricity is not obviously influenced by flybys.  Extreme outcomes (inclination and eccentricity) can be explained by planet-planet scattering alone. The median eccentricity is affected by the stellar birth cluster, with a profile that follows roughly
$\max(0.4 \exp(-x/1.1),0.024)$ (shown in red). Broadly, the eccentricity results are consistent with the results of \cite{heggie_rasio_1996_apj_282}, who found that the change in the eccentricity of a binary due to a distant encounter transitions to an exponential form for sufficiently small $\qf$.  A detailed comparison is difficult to make because the systems studied here all are multi-planetary systems, with the base level of eccentricity excitation higher than what is expected for most flybys.  One noticeable difference between our results and those of Heggie \& Rasio is that the maximum eccentricity remains below 0.5.  This may be due to ejections of highly excited planets.  Both the eccentricy and inclination profiles will be discussed further in Section \ref{sec:excitation_discussion_100}.   In the next section, we change focus from dynamical heating to \ACBc{major changes in the orbits of the planets} as a result of scattering.

\subsection{Major Changes in Planetary Orbits}

\begin{figure}
\begin{center}
\includegraphics[width=8.5cm]{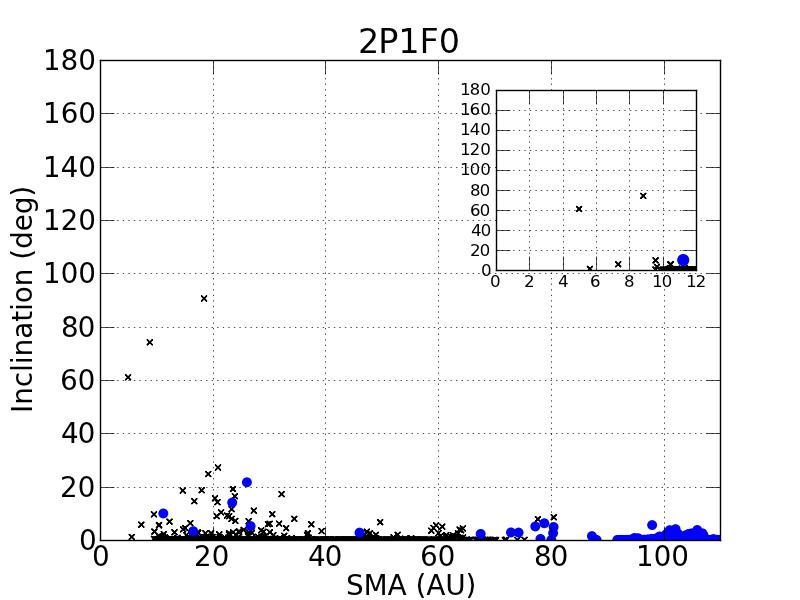}\includegraphics[width=8.5cm]{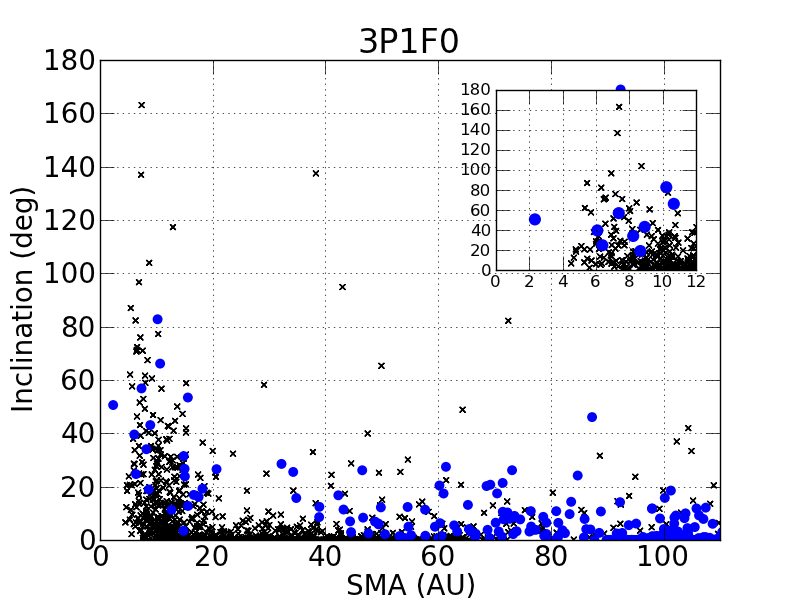}
\includegraphics[width=8.5cm]{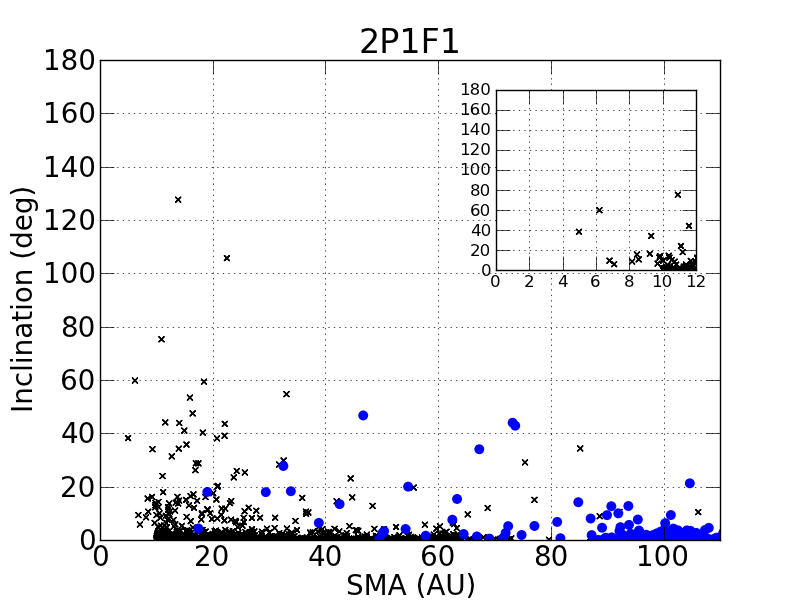}\includegraphics[width=8.5cm]{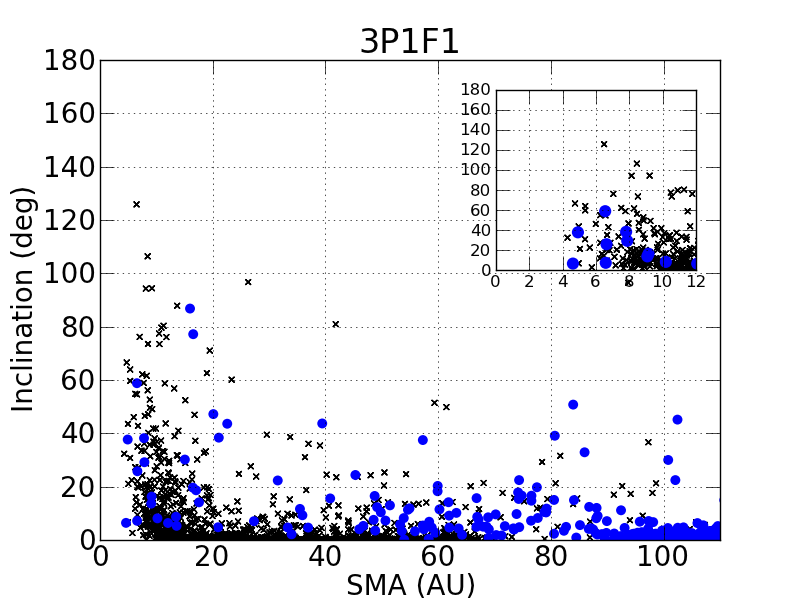}
\caption{Semi-major axes (SMA) versus inclination for each planet at the end of the simulation.  Black crosses represent all planets that were initially interior to the 100 AU wide-orbit planet (blue).  Scattering usually places inner planets on wide orbits, but can also place wide-orbit planets onto orbits of a few AU.  }
\label{fig:a_i}
\end{center}
\end{figure}

Planet-planet scattering, with and without close encounters, can lead to \ACBc{radial} mixing of planetary orbits, bringing outer planets inward by several orders of magnitude and placing inner planets on very wide orbits.  In Figure \ref{fig:a_i} we show the distribution of planet semi-major axes and  inclinations relative to the $x-y$ plane at the end of each simulation.  Black crosses represent planets that were originally interior to the 100 AU planet, and blue circles represent the planets that were initially at 100 AU.  Because the simulations with flybys have three times the number of systems as the simulations without flybys, we randomly select a third of the systems in 2P1F1 and 3P1F1 to show on the plots.   Planets that are initially on wide-orbits can be scattered to a semi-major axis that is interior to the initial innermost planet.  Flybys do increase the amount of this \ACBc{radial transport}, but planet-planet scattering alone will lead to large scale mixing.  It should thus be stressed that {\it observing a planet at a given location in a disk is not by itself indicative of how and where the planet formed.}  This will be addressed again in Section 5.4. 

%\MJPa{You could link directly to Fig 14 from here (that extreme outcome is NOT addressed in \S 4.3 but rather is delayed to \S 5.4, so it might be worth making the link obvious)}

\begin{figure}
\begin{center}
\includegraphics[width=8.5cm]{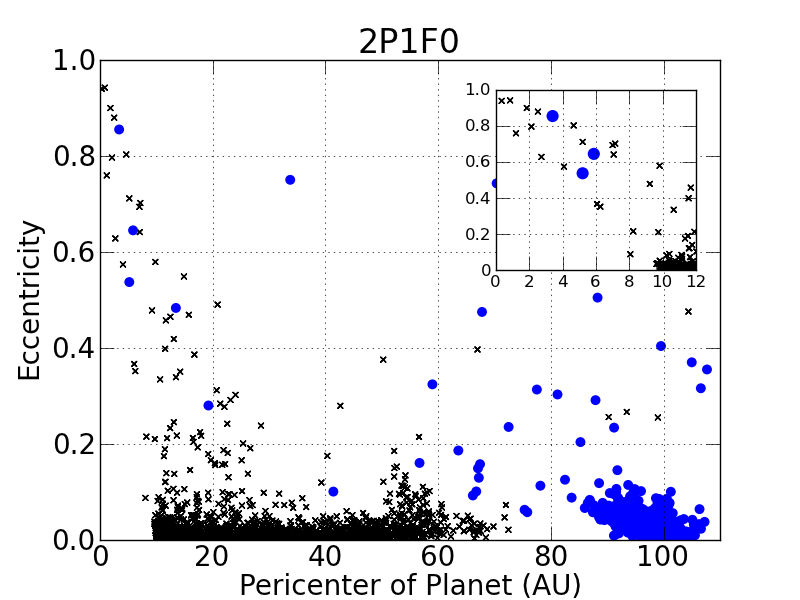}\includegraphics[width=8.5cm]{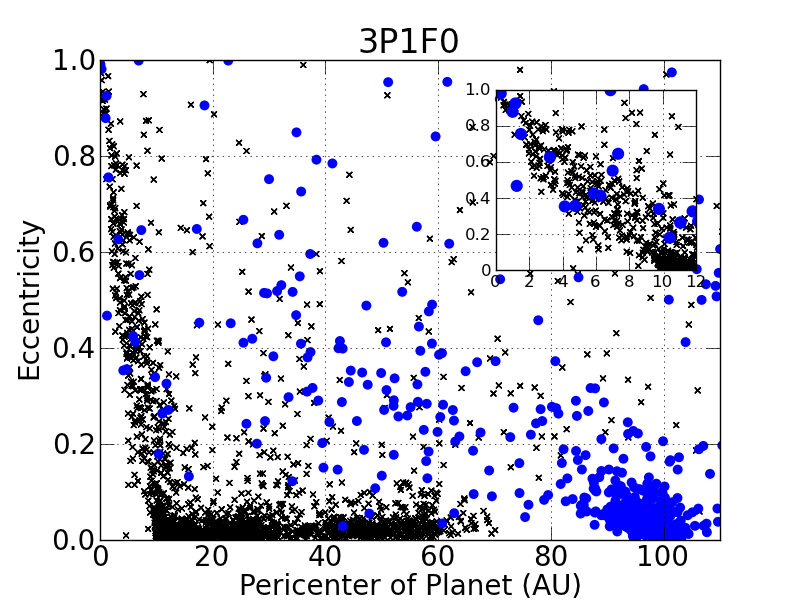}
\includegraphics[width=8.5cm]{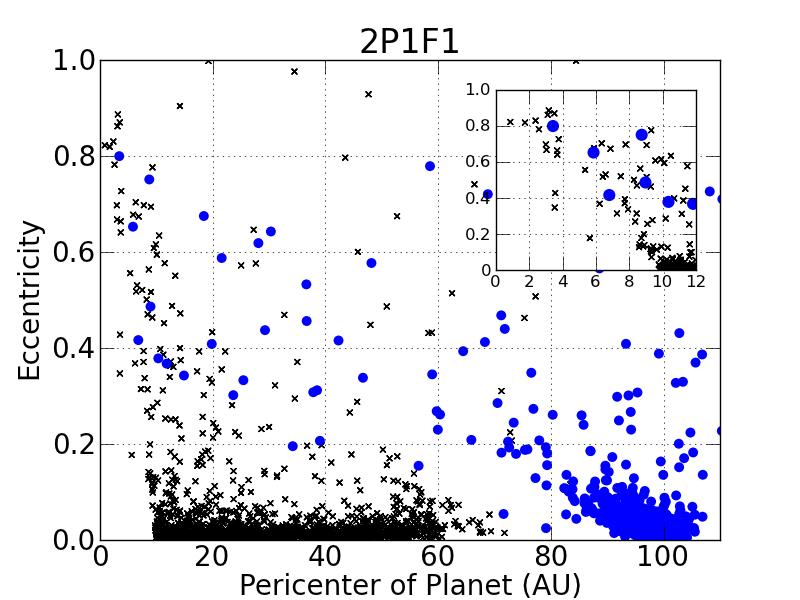}\includegraphics[width=8.5cm]{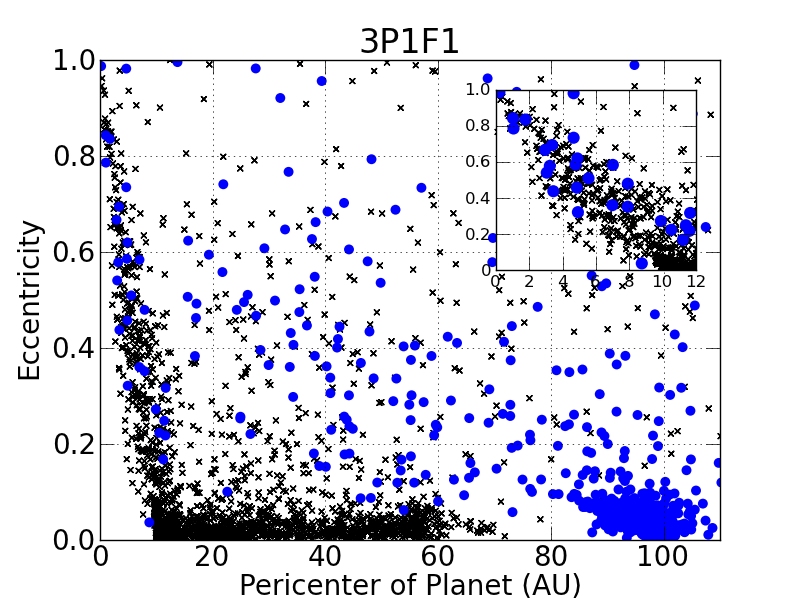}
\caption{Similar to Figure \ref{fig:a_i}, but for planet orbital eccentricity and pericenters.}
\label{fig:qplan_e}
\end{center}
\end{figure}

\begin{figure}
\begin{center}
\includegraphics[width=8.5cm]{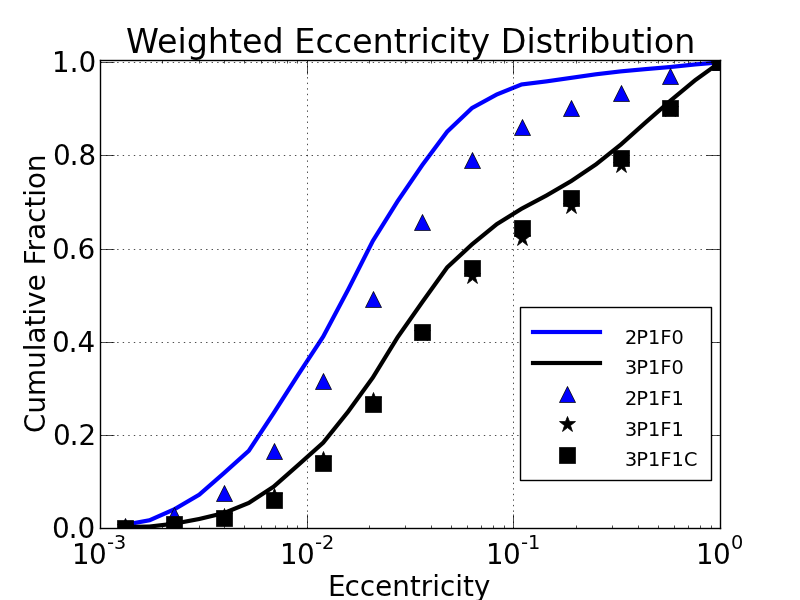}\includegraphics[width=8.5cm]{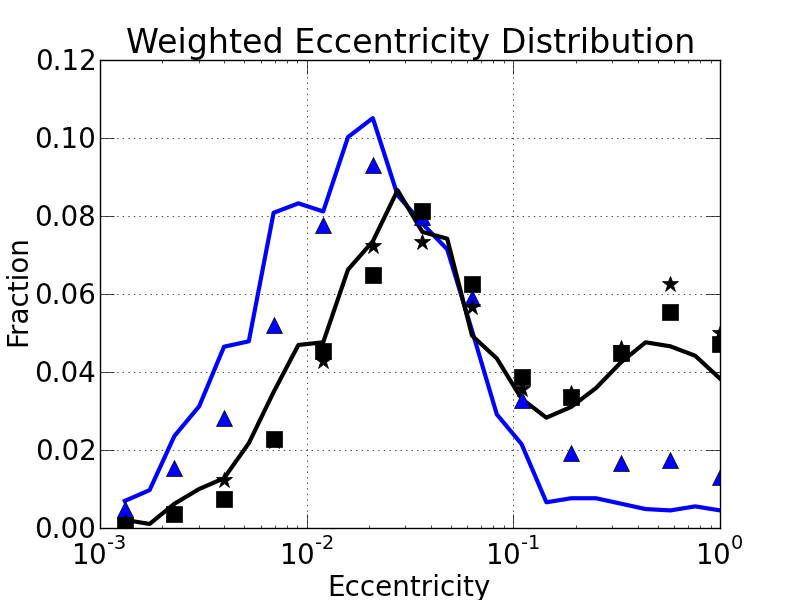}
\caption{Cumulative and specific distribution functions for the eccentricities of all planets weighted to account for a realistic distribution of $q$ for a cluster of $N=300$.  Perturbations by passing stars have a small effect on the eccentricity distribution of the planets, with planet-planet excitation clearly dominating the distribution function.}
\label{fig:hist_e}
\end{center}
\end{figure}

The connection between planet pericenters and eccentricity is shown in Figure \ref{fig:qplan_e}.  Most of the planets on small pericenters are the result of high-eccentricity orbits. Nevertheless, some planets do have small pericenters with eccentricities that are not near unity.  In particular, an initially wide-orbit planet  ($a=100$ AU) has a  $q\sim1$ AU and $e\sim0.5$ at end of the 3P1F0 simulation. The weighted eccentricity distribution is shown in Figure \ref{fig:hist_e}.   The median eccentricities are 0.015, 0.038, 0.019, and 0.047 for 2P1F0, 3P1F0, 2P1F1, and 3P1F1, respectively.  Flybys do influence the distributions, but the dominant effect for producing large eccentricities remains the number of planets/planetary orbital density of the system.

 The large degree of \ACBc{radial} mixing seen in these simulations, with and without flybys, gives rise to a small but non-negligible population of extreme systems.  We explore these outcomes in the next section.

%\begin{figure}
%\begin{center}
%\includegraphics[width=8.5cm]{data/hist_a_all_renorm.png}\includegraphics[width=8.5cm]{data/hist_a_gi_renorm.png}
%\caption{Normalized cumulative distributions for the semi-major axes of all planets (left) and only the wide-orbit planets (right).  Scattering and stellar flybys broaden the distribution of wide-orbit planets, but well over 50\% of these planets remain at 100 AU. For comparison, the initial distribution for 3P1F1 is shown (IC).}
%\label{fig:hist_a}
%\end{center}
%\end{figure}

\subsection{Extreme Outcomes\label{sec:extreme_outcomes}}

\begin{figure}
\begin{center}
\includegraphics[width=8.5cm]{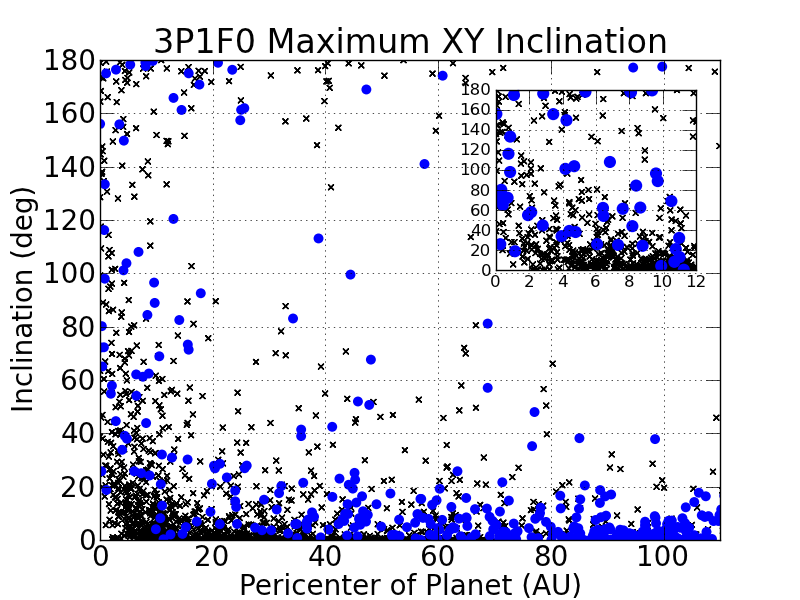}\includegraphics[width=8.5cm]{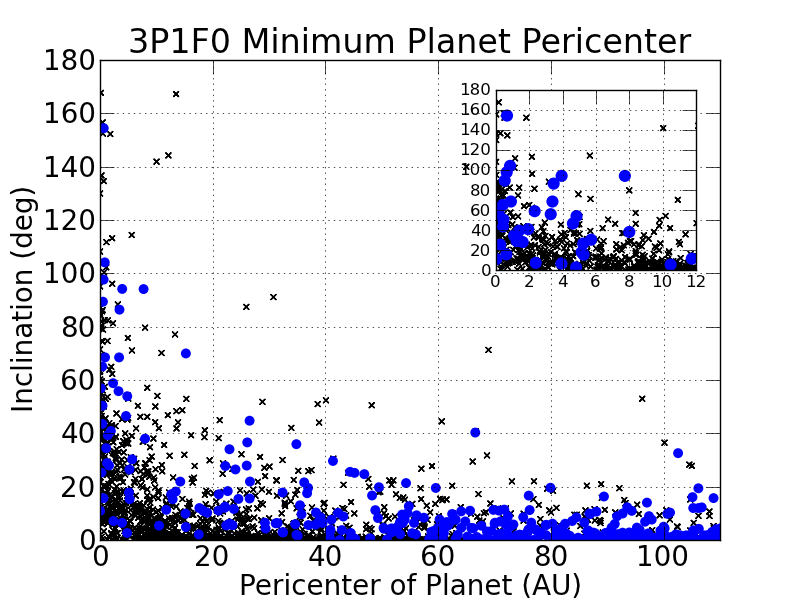}
\includegraphics[width=8.5cm]{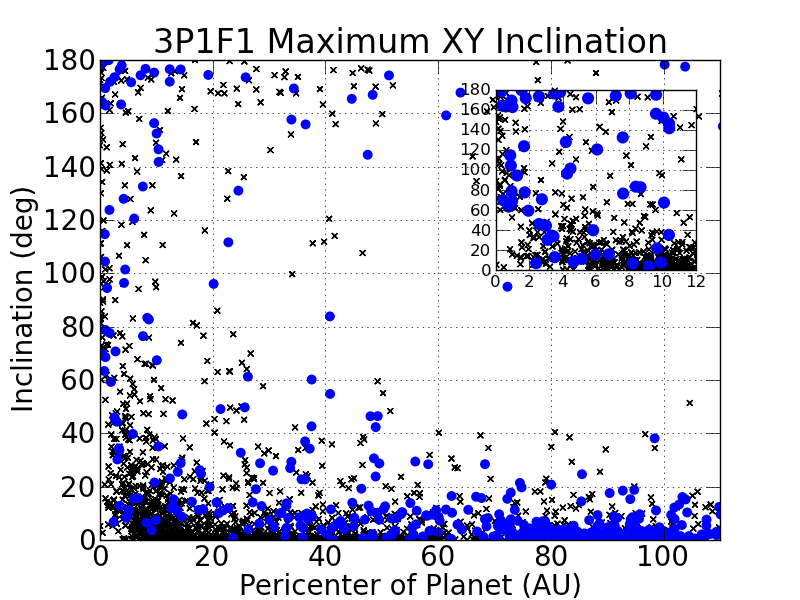}\includegraphics[width=8.5cm]{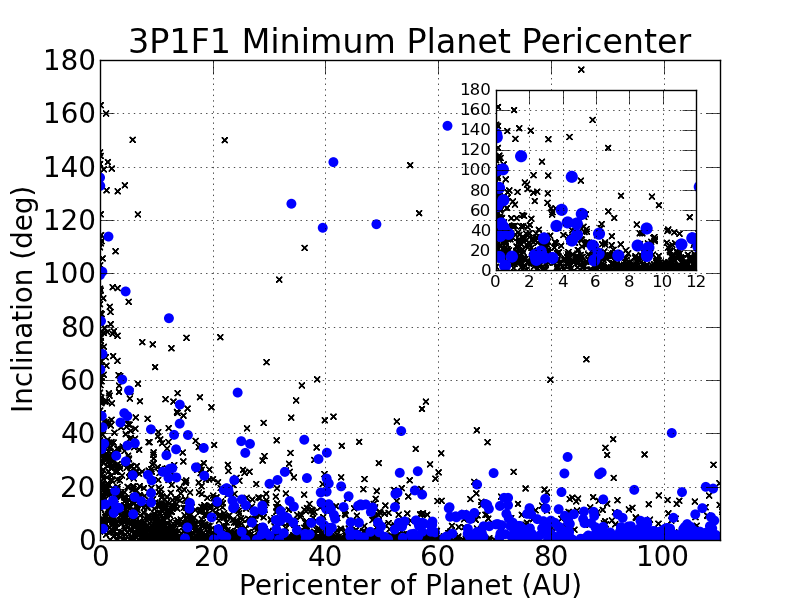}
\caption{Pericenter and  inclination relative to the $x-y$ plane at the time when the host system has any one planet reach the maximum inclination that ever occurs during its evolution, as well as the pericenter and inclination at the time when the system has a planet reach the minimum pericenter that ever occurs. Black crosses represent planets with initial positions inside the 100 AU wide-orbit planet (blue circles). The star's spin is envisaged in these simulations to be normal to the $x-y$ plane.  There is a pileup of planets on retrograde orbits for the maximum inclination plots.  Many of these planets are in the process of being ejected. }
\label{fig:maxever_scatter}
\end{center}
\end{figure}

In Figure \ref{fig:maxever_scatter}, we show each planet's pericenter and  inclination relative to the $x-y$ plane  at the time when the host system has any one planet reach the maximum inclination that ever occurs during the system's evolution, as well as at the time when the system has a planet reach the minimum pericenter that ever occurs. Only 3P1F0 and 3P1F1 are shown, and all planets in a given system are plotted.  Parameter space is filled much more evenly than what is seen from the end state of each system only, with many wide-orbit planets spending time in the inner nebula or on highly inclined orbits. Many of the planets on retrograde orbits in the maximum inclination plot represent snapshots just before they become ejected.  \ACBc{We emphasize that this does not represent the ``end state'' of the system, arbitrarily defined here as $10^8$ yr. The purpose of selecting the systems at times of minimum pericenter and maximum inclination is to demonstrate that planets in any given planetary system could have occupied a much larger fraction of the disk than inferred from the end state. Furthermore,  if some of these planets enter within 0.1 AU, then their evolution might be altered by tidal damping  (the effects of which are {\it not} included in the plotted simulation ensembles),} possibly locking the inclination of the planet into place while its semi-major axis is reduced \citep[see also Section \ref{sec:hotjupiters}][Payne et al. 2012, in prep]{nagasawa_etal_2008_apj_678}.  
%The fraction of systems with a planet that has  penetrated $q<0.25$ AU is 0.005, 0.05, 0.01, 0.06 for 2P1F0, 3P1F0, 2P1F1, and 3P1F1, respectively. 
The fraction of systems with a planet that has  penetrated $q<0.1$ AU is 0.003, 0.03, 0.009, 0.04 for 2P1F0, 3P1F0, 2P1F1, and 3P1F1, respectively. 
 Figure \ref{fig:hist_q0p1} shows unweighted cumulative distributions of the inclination  relative to the $x-y$  plane for all planets that have a minimum pericenter $q<0.1$ AU.      We  use the inclination relative to the  $x-y$  plane because we are interested in the orbital-spin alignment between the planet and star for RM measurements.  The unweighted distributions are shown because the fraction of planets is too small to apply weights with a reasonable degree of confidence. Nonetheless, the 3P1F0 simulation will serve as a baseline. In 3P1F0 and 3P1F1, the median inclination for planets that have $\qf <0.1$ AU is between 40 and 60 degrees.  Very small pericenters do occur for the 2P1F0 and 2P1F1 simulations, but these are rarer and are much more influenced by flybys than the 3P1F1 simulation.  Flybys do little to change the distributions for the 3P1F1 simulation, but have very strong consequences for sparsely populated planetary systems. 
 
Figure \ref{fig:maxever_scatter} shows the initial masses and semi-major axes for all planets that are included in the minimum pericenter distribution.  Planets with a minimum $q<0.1$ AU come from a range of locations, including very wide orbits,  and have a range of masses.

%\begin{figure}
%\begin{center}
%\includegraphics[width=8.5cm]{data/2p1p0_mixing_a10.png}\includegraphics[width=8.5cm]{data/2p1p1_mixing_a10.png}
%\includegraphics[width=8.5cm]{data/3p1p0_mixing_a10.png}\includegraphics[width=8.5cm]{data/3p1p1_mixing_a10.png}
%\caption{Planet semi-major axes and masses at initial values for all planets that orbit within $a=10$ AU at the end of the simulation.}
%\label{fig:mixing_maxa_ic}
%\end{center}
%\end{figure}

The large degree of \ACBc{radial} mixing that arises from multiple scattering events in the same system can result in planet-planet collisions, even at these large separations.  \ACBc{We define a merger event in the Mercury code to occur whenever planets pass within 2 $R_J$ of each other.  This gives an optimistic limit, as some of such collisions could only be hit and runs}.   In some cases, this places the planet over the deuterium burning threshold, assuming almost all of the combined planets mass is retained. 
%Energy conservation in these cases have been checked, and are better than $10^{-8}$. 
The frequency of such collisions is $\lesssim 1$\%. The afterglows of such collisions may be observable, and in at least one case, may have already been observed \citep{mamajek_meyer_2007_apjl_668,miller-ricci_etal_2009_apj_704}. In addition, debris trails may be a relic of collisions long after the afterglow has faded \citep{wyatt_dent_2002_mnras_334}.  

%\begin{figure}
%\begin{center}
%\includegraphics[width=8.5cm]{data/2p1p0_m_a_all.png}\includegraphics[width=8.5cm]{data/2p1p1_m_a_all.png}
%\includegraphics[width=8.5cm]{data/3p1p0_m_a_all.png}\includegraphics[width=8.5cm]{data/3p1p1_m_a_all.png}
%\caption{Planet semi-major axes and masses for all surviving planets at the end of the simulation. Although rare, planet-planet collisions have lead to the growth of some planets to masses $M>10 M_J$.  Black crosses represent planets with initial positions inside the 100 AU wide-orbit planet (blue circles). }
%\label{fig:collisions}
%\end{center}
%\end{figure}

Another extreme outcome of planet-planet interactions is ejection of a planet from the system.  Figure \ref{fig:ejection_ratio} shows the fraction of systems per $\qf$ bin for which all planets remain bound to their system at the end of the simulations. The bin width is allowed to vary to ensure that there are equal systems per bin. The final bin that does not make the cutoff is ignored. For comparison, the fraction of all systems that have kept all of their planets is 0.93 and 0.47 for 2P1F0 and 3P1F0, respectively. Flybys are directly responsible for ejections for $\qf\lesssim 2.5$ times the radial extent of the planetary system, $R_{\rm sys}$, which is consistent with previous work \citep[e.g.,][]{adams_laughlin_2001_apj_150}. No significant wing is seen for $\qf > 250$ AU, demonstrating that  if flybys that are greater than 2.5 $R_{\rm sys}$ contribute to planetary ejections,  the effect can be at most a few percent of the total fraction of systems that are destabilized. At first, these results may seem at odds with those of \citet{malmberg_etal_2011_mnras_411}, who find that flybys lead to a significant decrease in the long-term stability ($10^8$ yr) of a system, increasing the the fraction of systems that have had at least one ejection by factors $\sim 3$-9, depending on the mass distribution of planets. However, their measurements are taken for flybys inside $\qf < 3.3 R_{\rm sys}$.  By this measure, we would also conclude that flybys are an important ejection impetus, so the results are consistent. However, we will show in Section \ref{sec:excitation_discussion_general} that the frequency of flyby-induced ejections for field stars is limited to a few to 10\%. The formation mechanism for free-floating planets is primarily auto-ionization. 

Each system that experiences ejections can produce multiple free-floating planets.  Using the results without applying any weightings, we find that about 0.08, 0.21, 1.1, and 1.2 free-floating planets per system are generated for 2P1F0, 2P1F1, 3P1F0, and 3P1F1, respectively.  While flybys have a strong effect on sparsely populated planetary systems, having four instead of three planets in the system is more important for contributing to multiple ejections, as there are more planets in the system that are able to decay \citep{juric_tremaine_2008_apj_686}.  The scenarios explored here do not produce $\sim 2$  free-floating planets per system, as suggested to exist by \citep{sumi_etal_2011_nature_473}.  However, these microlensing observations only probe separations $\gtrsim 10$ AU, and not strictly whether the planets are unbound.  In this case, there is no contradiction between the observations and our results.  It is also possible that systems have initially more giant planets than envisaged in these simulations.  Most of the 2P1F0 systems have only one ejection, leaving a two-planet system.  For 3P1F0, two planets are ejected, leaving a two-planet system. If we extrapolate this rate and assume that half of all systems experience an ejection cascade, six-planet gas giant systems would need be to be a common formation scenario for the microlensing results to correspond to true, \ACBc{ free-floaters \citep[although see][for a more detailed estimate]{veras_2012_MNRAS_421L}.} 

\section{Discussion\label{sec:discussion}}

\subsection{Excitation due to Stellar Flybys: $R_{\rm sys}=100$, 30 AU\label{sec:excitation_discussion_100}}

Figure \ref{fig:median_by_pericenter} demonstrates that the maximum mutual inclination that we can expect for a given system is most sensitive to the number of planets in the system, and not to the closest approach of a stellar flyby.  In contrast, it also demonstrates that the median mutual inclination among planets in a given system is sensitive the pericenter of a stellar flyby, and not strongly to the number of planets in a system.  We can use this result to estimate the median mutual inclination for planets in planetary systems that come from a range of initial cluster sizes.  We find that $i_{\rm median}\approx\max(25\exp(-x )+2.15/(0.1+x),0.1)$ degrees, where $x=q/R_{\rm sys}$ degrees, where $R_{\rm sys}$ is the system's initial radial extent. The floor of 0.1 degrees takes into account the base-level inclination excitation from planet-planet scattering alone, which we take to be the average of 2P1F0 and 3P1F0 from Table \ref{table:median_inclinations}.  In the simulations presented here, $R_{\rm sys}=100$ AU. The expectation median inclination for a distribution is
\begin{equation}
\left< i_{\rm median}\right>=\int_{N_0}^{N_1} \frac{d\xi_N}{d N} \left(\int_{0}^{q_{\rm large}}\frac{\partial \xi} {\partial q}~i_{\rm median}dq \right) dN / \eta'.\label{eqn:expected_i}
\end{equation}
We restrict the number of encounters to be less than unity for a given $N$ and $q$ because we do not want to add extra weight to very distant encounters. As such, $q$ is only integrated until every system on average has one encounter ($q_{\rm large}$). A similar approach can be taken for the median eccentricity in systems by replacing $i_{\rm median}$ with $e_{\rm median}\approx \max(0.4 \exp(-x/1.1),0.026)$.  We set the basal eccentricity to 0.026, which is the average of the median eccentricities for the 2P1F0 and 3P1F0.   The results are shown in Table \ref{table:expectation_values}. The median mutual inclination for a distribution of planetary systems that had a planet at 100 AU initially will be between about two and five degrees. Likewise, the median eccentricity will be between 0.04 and 0.08, depending on the cluster size. 
% If we take the velocity dispersions for the radian and azimuthal directions to be $\delta v_r= e v_K$ and $\delta v_z = i v_K$, we find that these components are roughly equally partitioned.
The relationships for $i_{\rm median}$ and $e_{\rm median}$ depend on $R_{\rm sys}$.  If we take $R_{\rm sys}=30$ AU, analogous to the Solar System's size for the major planets, then the expected median inclination is only $\sim 1$ degree and the $\left<e_{\rm median}\right>$ is between 0.03 and 0.04.  
%In this case, the radial velocity dispersion is $\sim $1 ($N_0=10$) to $\sim 3$ ($N_0=100$) times the vertical dispersion. 

We also use our results to explore the frequency of ejections. From Figure \ref{fig:ejection_ratio}, flybys trigger additional ejections for $\qf/R_{\rm sys}\lesssim2.5$.  The ejection distribution is sensitive to the number of planets, with the frequency of ejections about 8 times larger in 3P1F0 than in 2P1F0.  As with the other profiles, we take the average between these basal levels.  The functional form for the fraction of systems that retain all of their initial planets is then $f_{\rm survive}= 0.7(\qf/R_{\rm sys})$ for $\qf/R_{\rm sys}<2.5$ and 0.7 otherwise. Table \ref{table:system_sizes} gives the results.  Flybys have a 4-15\% effect on systems with $R_{\rm sys}=100$ AU and, based on scaling, we predict a 1-4\% effect on systems with $R_{\rm sys}=30$ AU.

The flyby rate {\it and} the distribution of cross sections of planetary systems is a determining factor for the efficacy of flybys on shaping planetary system architectures.  In the next section, we  place constraints on this distribution and  build on the results from the above $R_{\rm sys}=100$ and 30 AU cases.

\subsection{Excitation due to Stellar Flybys: Integration Over Expected $R_{\rm sys}$ \label{sec:excitation_discussion_general}}

The fraction of highly inclined planets that are produced by flybys is dependent on the \ACBc{effective} cross section of the planetary system. 
\ACBc{In general, the effective cross section is not necessarily equivalent to the geometric cross section, but because we are assuming that the orbits for the outermost planets are initially circular, we take the cross section to be $\pi a_{\rm outer}^2$ , where $a_{\rm outer}$ is the outermost planet's semi-major axis.}  
While the distribution of planetary sizes is at this time unknown, we can use the distribution of specific angular momentum among low-mass cloud cores as a proxy. 
Tables 3, 4, and 5 from \citet{caselli_etal_2002_apj_572} provide data for 20 sources that have enough information for estimating a given core's angular momentum, $J$ \ACBc{ 
\citep[see also, e.g.,][for additional sources and discussion regarding the internal kinematics of cloud cores]{shu_etal_1987_ARAA_25,myers_benson_1983_ApJ_266,goodman_1993_ApJ_406,barranco_goodman_1998_ApJ}.  } 
We convert from the Caselli et al.~derived velocity gradient values $\cal{G}$ to $J$ by assuming virial equilibrium and constant density cores, which allows us to write $J\approx 0.4 R_{\rm core}^2 (4.86\times10^2 \mathcal{G}/n_{\rm vir})^{1/2}$ \citep[see, e.g.,][]{goodman_1993_ApJ_406}. 
Here $\mathcal{G}$ is in km s$^{-1}$ pc$^{-1}$ and the average number density of the core $n_{\rm vir}$ is in cm$^{-3}$.  
Based on these assumptions, the {\it initial} size of the protoplanetary disk is $R_J\approx J^2/(GM_{\rm vir})$, i.e., the radius we expect collapse to be halted by the system's angular momentum.  
In this estimate, we use the virial cloud core mass $M_{\rm vir}$ as derived by Caselli et al.  
As shown in Table \ref{table:system_sizes}, about 30\% of systems could possibly form a substellar companion at semi-major axes 80 AU or greater.  
This estimate is admittedly crude, but allows us to extend our estimates from section \ref{sec:excitation_discussion_100} by accounting for disk sizes.   
We repeat the calculations in the previous section, but for $R_{\rm sys}=20$, 60, 100, 140 and 180 AU, where the result for each bin is weighted by the corresponding fraction. Under these assumptions, $\left<i_{\rm median}\right>\sim1$ to 3 degrees, $\left<e_{\rm median} \right>\sim 0.03$ to 0.06, and $f_{\rm survive}\sim 0.6$ to 0.7. 
At least for the systems studied here, most massive planets remain bound to their host star, with about 30-40\% of systems generating free-floating planets.  
We conclude that planet formation and subsequent planet-planet interactions, not stellar encounters, determine the final eccentricities and inclinations of the typical planetary system. 

 \ACBc{If, on average, the outermost planet in a system forms at a fraction of $R_{\rm J}$, then the effect of flybys is further marginalized than what we find here. Even if planets form at $R_{\rm J}$, they will not necessary stay there in a rapidly evolving, young, massive disk \citep{baruteau_etal_2011_mnras_416,michael_etal_2011_apj_737,zhu_etal_2012_ApJ_746}. This would reduce planetary system sizes and again the effects of flybys would be further marginalized.  In contrast, it may be possible to move massive planets outward \citep{crida_etal_2009_apjl_705} as disks transition away from their initial, massive state. This would increase the influence of flybys. The relative importance of these effects at this time is unknown, so although only approximate, we find our above calculation to be reasonable with the information available.}

\subsection{Observationally Constraining Formation Scenarios for Wide-Orbit Planets\label{sec:constrain_wideorbits}}

The formation mechanism of wide-orbit planets can potentially leave distinct dynamical signatures.   First, consider the fraction of planets with semi-major axes between 80 and 200 AU.  If we assume that every disk forms a planet {\it in situ} at its nascent cloud's angular momentum barrier and we further assume that this planet does not migrate, then about 18\% of systems should have a planet with a semi-major axis in the chosen annulus (30\% of all systems  with a survival rate of about 60\%) .  We do note that even if planets form  at large radii, they will not necessarily stay at this location \citep{baruteau_etal_2011_mnras_416,michael_etal_2011_apj_737}.  Nonetheless, the rate of occurrence may be much larger than the 3\% that is found in scattering simulations \citep[][see their Fig.~4]{veras_etal_2009_apj_696}. 

If migration or the rate of planet formation on wide-orbits produces a frequency that is indistinguishable from pure scattering, we still expect to observe significant differences in the orbital distributions between {\it in situ} formation and scattering only.  From the scattering experiments of  \citet{veras_etal_2009_apj_696}, planets with semi-major axes between $\sim100$ and 200 AU have eccentricities that are roughly evenly distributed between 0.4 and 1.0, with the lower bound increasing with $a$ such that the lower limit on the eccentricity is about 0.8 by $a\sim 1000$ AU. In contrast, the median eccentricity for the outermost planet in the simulations presented here is $\sim0.13$, where we have averaged the median eccentricities for $a\sim100$ and 200 AU.  These eccentricity differences lead to highly distinguishable features in observable distributions.  

We explore the effect that different eccentricity distributions have on observables by making the following assumptions and cuts in parameter space: We limit the semi-major axes of the planets to be between 100 and 200 AU.  In this regime, the planet semi-major axis distribution due to planet-planet scattering, based on the simulations of  \citet{veras_etal_2009_apj_696}, is $dN_{\rm pl}/da\propto a^{-2.75}$, where the power law profile is taken directly from their simulation data.  The $a$ distribution for {\it in situ} formation is not so cleanly defined from core velocity gradients, so we adopt the same profile for the following comparisons.  The underlying $a$ distribution will have an effect on the profiles, but strong features are insensitive to this assumption. We produce distributions of planet radial separations from the star and of star-planet relative radial and azimuthal proper motions by using Keplerian orbits with a range of eccentricities and semi-major axes.     Each orbit's contribution to the distribution is weighted by  $a^{-2.75}$ and by the corresponding inverse orbital period.  The inverse period weighting is necessary to keep planets on longer orbits from having extra weight, as only the fraction of the planet's orbit that is within the chosen annulus matters.  Radial and tangential velocity components are transformed into proper motions relative the host star to highlight observable constraints, with the tangential proper motion given as a deviation from a circular orbit at the given location.  We assume a distance of 10 pc, and the distributions are assumed to be face-on.

Figure \ref{fig:obs_dist_pm} shows the distributions of the radial separation, $R$, the radial proper motion, $V_R$, and the deviation of the tangential proper motion, $\vert V_{\rm phi}-V_{c}\vert$,  for $e=0.13$,  $e=0.40$, and an integration over an even distribution of eccentricity in the range $[0.4, 0.95]$.  Proper motions are strongly peaked near $V_R\sim \left(\mu/a({\rm outer})\right)^{1/2} e$ and $\vert V_{\phi}-V_c\vert \sim \left(\mu/a({\rm outer})\right)^{1/2} e/2$, where $a({\rm outer})$ is the largest semi-major axis included in the distribution (200 AU here) and $\mu=G(m_{\rm star}+m_{\rm planet})$. The radial distribution goes as $r^{-2.75}$ in regions where the pericenter and apocenter are both within the 100-200 AU annulus, which is expected from the assume semi-major axis profile.  The peak in the proper motion distributions  for the range of $e$ is broadened by high eccentricities, with a tail extending to large $e$, but remains near the distribution with only  $e=0.4$ because the lower eccentricity planets are more likely to be seen within a fixed-width annulus.   The radial distributions between a fixed $e$ and the range of $e$ have very different shapes.

A single component distribution of the proper motion can strongly distinguish between scattering only and a significant {\it in situ} formation population, even if the underlying semi-major axis distribution is unknown.  Pure scattering will exhibit a peak in the $V_R$ distribution that is at proper motions that are three times larger than a peak that corresponds to only {\it in situ} formation on initially circular orbits.   As a result,  a limited search may be able to test whether a large component of wide-orbit {\it in situ} formation planets exist.  The model distributions cannot be used alone to constrain the formation mechanism of any single planet, but can be used to comment on how typical a given system is under the premise of the model. As an example, consider Fomalhaut b.  Based on \citet{kalas_etal_2008_science_322}, the proper motion of the planet candidate is about 100 mas/yr, largely North and in the projected tangential direction.  Taking this at face value, the tangential proper motion deviation is about 7 mas/yr for $R=119$ AU, which would correspond to $e=0.13$ for Fomalhaut b's motion to lie on the most probable deviation of tangential proper motion.  This eccentricity is the same as that already derived by Kalas et al., and is at the median value for {\it in situ} planet formation based on our scattering experiments.

\subsection{From Wide-Orbits to Hot Jupiters\label{sec:hotjupiters}}

\begin{figure}
\begin{center}
\includegraphics[width=8.5cm]{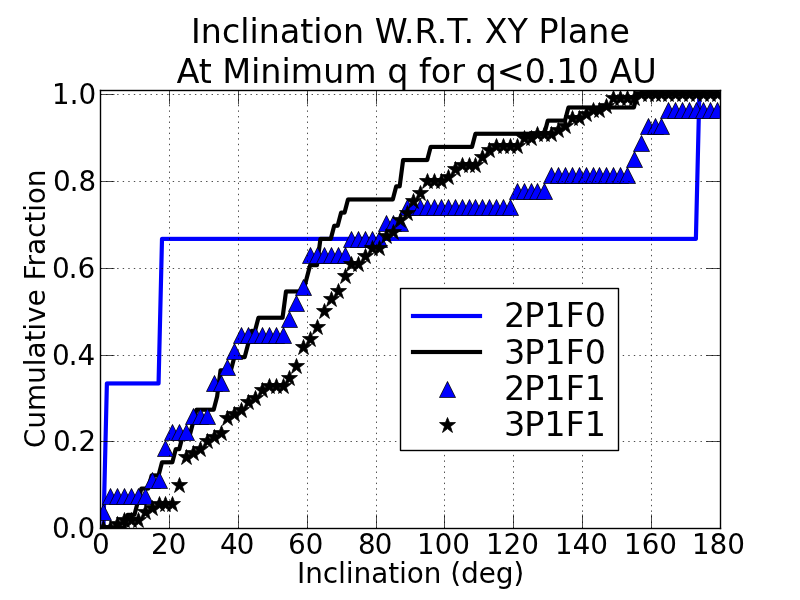}\includegraphics[width=8.5cm]{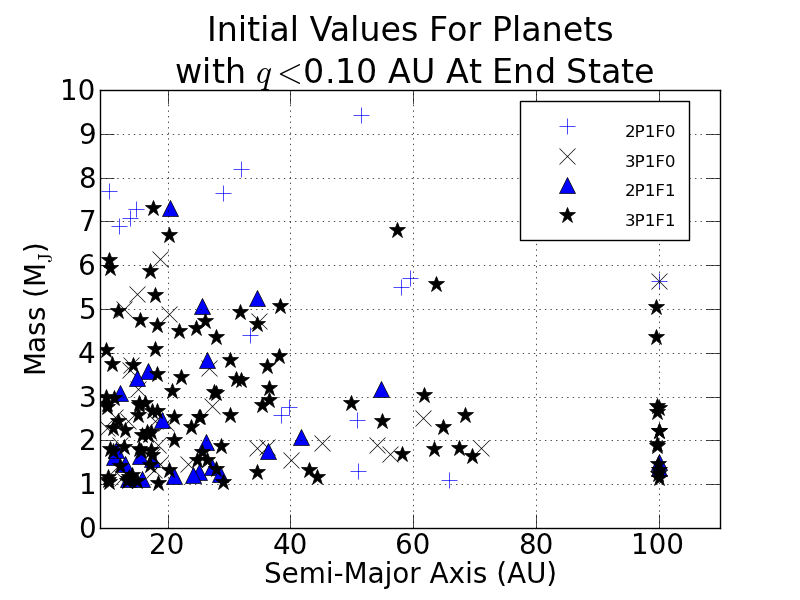}
\caption{Left: Cumulative distributions (unweighted) for all planets that have a pericenter $q<0.1$ AU at any time during the simulation. Right: Initial planet semi-major axes and masses for all planets that orbit within $q=0.1$ AU at some point during the simulations. Lower mass planets are preferentially scattered onto the more highly eccentric orbits.   }
\label{fig:hist_q0p1}
\end{center}
\end{figure}

\begin{figure}
\begin{center}
\includegraphics[width=15cm]{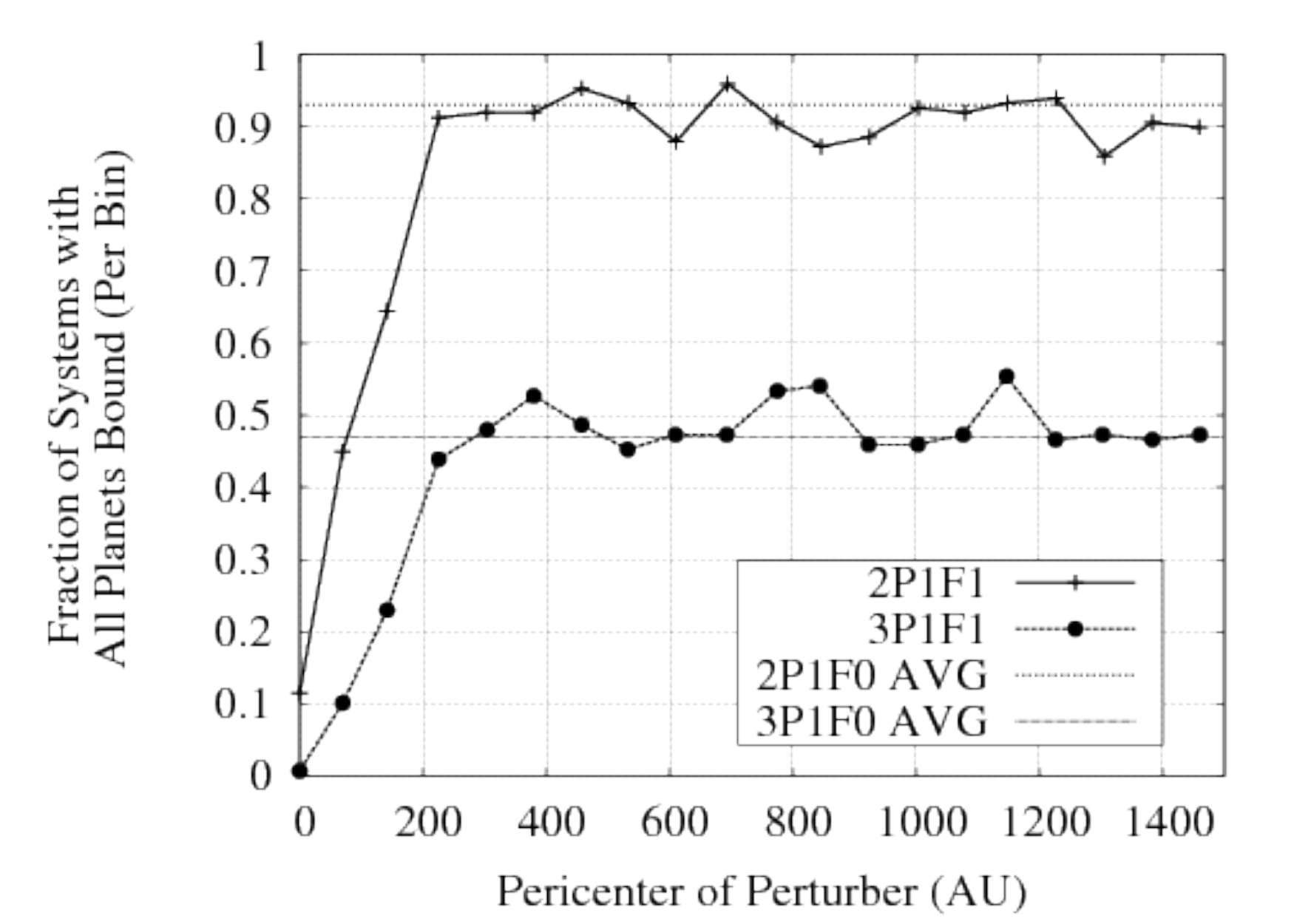}
\caption{The fraction of systems per $\qf$ bin for which all planets remain bound to their system at the end of the simulations. The bin width is allowed to vary to ensure that there are equal systems per bin (chosen here to be $\sim150$). The final bin that does not make the cutoff is ignored. For comparison with the fractions at large $\qf$, about 0.93 and 0.47 (lines) of all systems experience no ejections in 2P1F0 and 3P1F0, respectively.}
\label{fig:ejection_ratio}
\end{center}
\end{figure}

\begin{figure}
\begin{center}
\includegraphics[width=6cm]{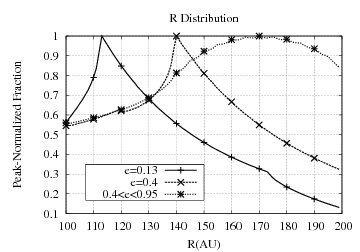}\includegraphics[width=6cm]{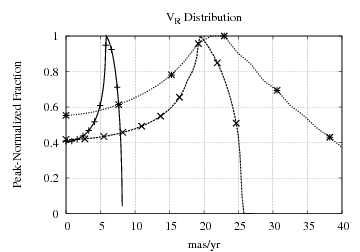}\includegraphics[width=6cm]{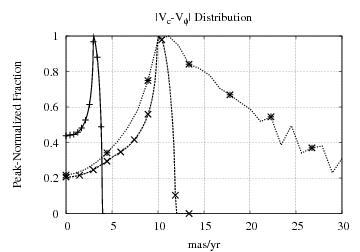}
\caption{ Radial separation (left), the relative radial proper motion (center) and the deviation of the relative tangential proper motion (right) distributions of planets on wide orbits for several eccentricity distributions. We assume a distance of 10 pc for conversion of the velocities to star-planet relative proper motions.\label{fig:obs_dist_pm}}
\end{center}
\end{figure}

\begin{figure}
\begin{center}
\includegraphics[width=8cm,angle=-90]{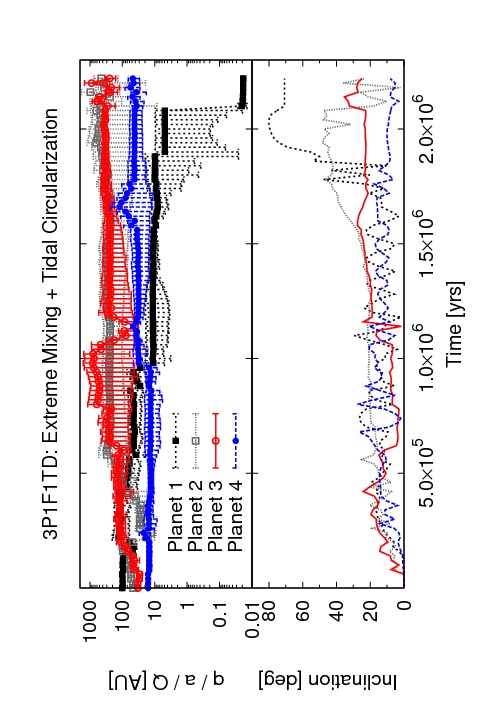}
\caption{Top: The pericenter $q$, semi-major axis $a$, and apocenter $Q$ for each planet in one of the 3P1F1TD systems.  Bottom: The evolution of the inclinations relative to the $x-y$ plane for the same planets shown in the top panel. The planets in the system become dynamically unstable, and the outermost planet becomes the innermost one.  Its orbit eventually becomes eccentric enough for the planet to pass by the star at $\sim0.02$ AU, where dynamical tides quickly circularize the orbit.  In this example, a hot Jupiter with an \ACBc{inclination relative to the $x-y$ plane $i\sim 70^{\circ}$, a semi-major axis $a=0.019$, and $e=0.013$} is made from a planet that  was initially at 100 AU.  }
\label{fig:scatter_out_to_in}
\end{center}
\end{figure}

In Section \ref{sec:extreme_outcomes}, we found that a small fraction of our systems have a planet with a pericenter $q<0.1$ AU  during some point of the system's evolution.  The fraction of systems with this outcome varies between tenths and several percent, depending on the total number of initial planets in the system and whether the system ever experiences a flyby.   We refer to these planets as hot Jupiter candidates, as many approach the star with a pericenter that is small enough for tidal friction to be important.   Some of these candidates are initially interior to the 100 AU planet, but some are initially on very wide wide-orbits (see Fig.~\ref{fig:hist_q0p1}).   When wide-orbit planets scatter close to the star, they do not do so directly.  Instead, they have multiple encounters with the inner planets, giving the planet a small pericenter.  To explore this possibility, we integrate 1000 systems from the 2P1F1 and 3P1F1 ensembles (500 each) using the same Mercury integration scheme that is used in the other ensembles, but with the inclusion of  a tidal damping model \citep{nagasawa_etal_2008_apj_678}.  We refer to these new ensembles as 2P1F1TD and 3P1F1TD to distinguish them from 2P1F1 and 3P1F1, which are run without tidal damping.   One system in 2P1F1TD and three in 3P1F1TD form a hot Jupiter from a planet that is originally at 100 AU.  While this is a small fraction of all planetary systems, it is  potentially a \ACBc{non-negligible} fraction of hot Jupiters.  Figure \ref{fig:scatter_out_to_in} shows an example of a hot Jupiter that forms from an inward scattering cascade of a planet that is initially at 100 AU. Note that the planet is also highly inclined, with an end inclination relative to the $x-y$ plane $i\sim 70^{\circ}$, \ACBc{a semi-major axis $a=0.019$, and an $e=0.013$}.  This emphasizes again that a planet's observed location alone is not indicative of where or by what mechanism the planet formed.

\ACBc{  The frequency of hot Jupiter candidates increases with decreasing planet mass (Fig.~\ref{fig:hist_q0p1}), and at least half are expected to have an inclination $i>40^{\circ}$ relative to the $x-y$ plane. The mass cutoff in our simulations is 1 M$_J$, but continuing this trend to the masses of the observed hot Jupiter population suggests that scattering should produce even more hot Jupiters at masses less than those studied here. Whether any given planet can ultimately become tidally captured will depend on the properties of that planet, with a strong, but nontrivial dependence on planetary radius at fixed mass  \citep[e.g.,][]{ivanov_papaloizou_2004_mnras_347, mardling_2007_mnras_382}.   If the initial radius of a planet is dependent its formation mechanism, then hot Jupiters might  be used a tool for exploring different modes of planet formation.  For example, if planets that form by direct instability have  higher specific entropy on average than core accretion planets of comparable composition and mass, then the radius evolution of these planets could be appreciably different \citep[e.g.,][]{spiegel_burrows_2011_arxiv}, which could have consequences for tidal capture of these objects.  }

\subsection{Outer Planets from Inner Planets\label{sec:out_from_in}}

\begin{figure}
\begin{center}
\includegraphics[width=14cm]{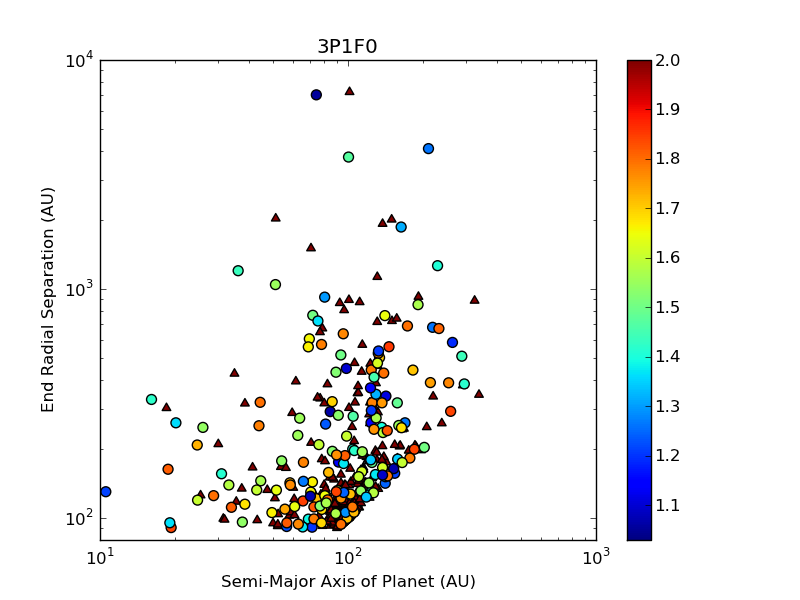}
\caption{A population of very wide orbit planets for 3P1F0, for which all planets with separations $>90$ AU at the end of the simulation are shown.  Circles represent planets that were not initially on wide orbits, while triangles represent planets that are initially at 100 AU.  The colorbar represents the log of the initial semi-major axis in AU.  Most of the triangles are clustered around $a,~r =100,$ 100 AU. For the planets with $r>1000$ AU, their orbits were interior to the 100 AU planet, but are preferentially at large $a$.  Several points do extend beyond the limits of this plot.  }
\label{fig:wide_orbit_scattered}
\end{center}
\end{figure}

\citet{veras_etal_2009_apj_696} have shown that a population of eccentric, wide-orbit planets can be produced by planet-planet scattering.  Their simulations included a high-density of planets between 3 and 7 AU.  In the simulations presented here, planets are already on moderate and wide orbits, which can give rise to very different scattered population.   First, consider the fraction of systems that have a planet at the end of the simulation with a radial separation from the star $>90$ AU.   For each ensemble, those fractions are 0.97, 0.92, 0.85, and 0.83 for 2P1F0, 2P1F1, 3P1F0, and 3P1F1, respectively, which demonstrates that at least a few to ten percent of systems with a planet initially  at 100 AU will not be observed to have one, even in the absence of flybys owing from planet-planet scattering and auto-ionization. If we exclude all planets that are initially on wide orbits, the fraction of wide-orbit planets is 0.016, 0.038, 0.17, and 0.20 for 2P1F0, 2P1F1, 3P1F0, and 3P1F1, respectively.  In many cases, the outermost planet is significantly farther out from the star at the end of the simulations than the outermost planet at the beginning of the simulations. Between about one and ten percent of systems with wide-orbit planets at the end of the simulation do not have the initial wide-orbit planet at such large separations.  Figure \ref{fig:wide_orbit_scattered} shows a population of very wide orbit planets for 3P1F0.  All planets that are bound and have a separation $>90$ AU at the end of the simulation are shown except for a small population that extend beyond the plot limits.  Circles represent planets that were not initially on wide orbits, while triangles represent the initial 100 AU planets.  The colorbar represents the log of the initial semi-major axis in AU.  Most of the triangles are clustered around $a,~r =100,$ 100 AU.  Planets on wide orbits can come from a very wide range of initial semi-major axes.  The widest planets $r>1000$ AU  are preferentially, but not entirely, the planets that began farther out in the disk. They are not, however,  preferentially the initial 100 AU planet.  Recall that 3P1F0 does not include flybys.  For planets that were not initially at 100 AU,  a population of  high pericenter, very long period planets is also possible.  In this context Sedna-like orbits can be produced by  a wide-orbit planet that is lost from the system.  Such a loss can be through ejection, or if planet formation can take place during the earliest stages of outer disk evolution, the scatterer could be a transient clump \citep{boley_etal_2011_apj_735}.

%155.277111021 1862.86218 3046.67773686

%401237 2 70.0450926731 519.449012 968.48247973

\subsection{Limitations of Current Study\label{sec:limitations}}

\ACBc{
The current study explores a single mass for the perturbing stars $(M_p=0.3 M_{\odot})$ and for the host stars of the planetary systems $(M_H=1.0 M_{\odot})$. 
In addition, it excludes encounters with binaries.  
Relaxing any one of these restrictions will have an effect on our results, which we discuss here.}

\ACBc{
First, the potential perturbation from a flyby on a planetary system scales with the mass of the perturber, so we expect more massive stars to lead to more ejections and to produce a larger median inclination among planetary systems than less massive perturbers.
Because we set $M_p$ to be near the median stellar mass, half of the flybys will lead to more energetic perturbations than captured in our simulations and half of the flybys will lead to less energetic perturbations.  
%Moreover, the mass range for $M_P<0.3 M_{\odot}$ is fairly compact, confined to about a factor of three, excluding brown dwarfs.   While the mass range for $M_P>0.3 M_{\odot}$ is much larger, the frequency of very massive stars becomes vanishingly small. 
We therefore do not expect the choice of perturber mass to be a major source of error in the study. }

\ACBc{
Second, the results presented here are most relevant to planetary systems around solar-type stars, resulting from our choice of $M_H$. 
Before the results can be extended to planetary systems around a distribution of host stars, we must consider the following:
(1) Low-mass host stars will have planets that can be more easily ionized by a perturber compared with higher-mass host stars, {\it ceterus paribus}.
(2) The specific angular momentum distribution of cloud cores may depend on cloud core mass, affecting the size distribution of planetary systems.
(3) Independent of angular momentum distributions, the formation of gas giants on wide orbits may depend on host star mass.
While, including a distribution for $M_H$ is necessary for understanding the role of flybys on planetary systems in general, it is difficult at this time to assess how our results will change for a realistic distribution of host star masses.}

\ACBc{
Third,  flybys by binaries can cause stronger perturbations to planetary orbits than flybys by single stars, owing to increased interaction cross sections and to resonant interactions between the planets and the binary whenever their periods are comparable.  Early studies of the multiplicity of {\it solar-type} stars suggest that about half of {\it solar-type} stars are in binaries
\citep{duquennoy_mayor_1991_AnA_248}, so it would therefore seem most relevant to focus on interactions between binaries and planetary systems instead of single stars and planetary systems as done here. However, recent studies show that multiplicity is a strong function of stellar mass \citep[e.g., see Fig.~12 of ][]{raghavan_etal_2010_ApJS_190}, with the implication that most  field stars ($\sim70$\%) are single \citep{lada_2006_ApJ_640L}.  In addition, only a fraction of binaries will have separations/periods that will significantly alter the interaction.  For example, both very short-period binaries and very long-period binaries will appear approximately as a single perturber.  To estimate the relevant fraction of binaries that will impact the planetary systems explored in this study, we integrate over the binary period distribution for solar-type stars given by \cite{raghavan_etal_2010_ApJS_190}. The fraction of binaries with  periods between 10 and 3000 yr ($\sim 5$ and 200 AU) is 0.41.  Combining this fraction with the frequency of binaries among all stars reveals that strong interactions with binaries, compared with single stars for the same closest approach, should be expected for $\sim 12\%$ of encounters.  Only a fraction of these encounters will have relevant close approaches to the planetary system, so the results we present here are not obviously affected by the exclusion of binaries.  }

\ACBc{
There are several additional caveats that should be mentioned.  (1) The binary period distribution is valid for solar-type stars in the field, while what is desired is multiplicity of stars while they are in their natal cluster.  (2) Observations suggest that the semi-major axis distribution of binaries is dependent on binary mass, with low-mass binaries being much more compact \citep{siegler_2005_ApJ_621}.  This will tend to decrease the influence of binaries compared with single stars by reducing the interaction cross sections of the binaries.  Finally, (3) we note that a very short-period binary will have a stronger perturbation on a planetary system than a single perturber on average by virtue of the binary being two stars.  This effect will give a slight skew of the effective median mass of perturbers to higher mass, but we do not expect this to be a major source of error in our results.}

\ACBc{
Overall, we do not find that the results of this study should be significantly affected by our choice to focus on single, $0.3\, M_{\odot}$ perturbers.}

\section{Conclusions\label{sec:conclusion}}

We have presented the results of a series of scattering experiments that investigate the dynamical outcomes for multi-planet systems that have planets initially on  wide-orbits.  The experiments compare  system architectures and scattering histories between  chiefly four ensembles: two of the ensembles do not include stellar flybys, while two include flybys by a 0.3 M$_{\odot}$ perturber.   All systems contain 1 planet at 100 AU.  Two of the ensembles are made of systems with two planets interior to the 100 AU planet, and two ensembles with three planets interior to the 100 AU planet. The four-planet systems fill the same semi-major axis range as the three-planet systems, so the four-planet systems are more densely packed.  The importance of flybys on these system architectures is then evaluated by direct comparisons between the non-flyby and flyby ensembles.   When possible, our results are rescaled for a distribution of planetary system sizes that are derived from literature values of cloud core velocity gradients and/or integrated over a range of natal cluster sizes for comparisons with field star populations.  We find the following key results:

\begin{enumerate}

\item High mutual inclinations within planetary systems are more likely to be due to planet-planet interactions than due to stellar flybys.  We find that for mutual $i>40^{\circ}$, flybys increase the fraction total fraction by about 1\%. Although a small increase overall, flybys can double the number of highly perturbed planets. 

\item Low mutual inclinations are strongly affected by stellar flybys.  Even if planets are born perfectly coplanar, the system's natal cluster will seed a substantial inclination dispersion.  We find the median inclinations for the three and four-planet systems at the end of the simulations to be about 0.24 and 0.86 degrees, respectively.  The same systems without flybys have a mutual inclination of about 0.08 and 0.12 degrees.  Figure \ref{fig:hist_i_weight} shows a clear separation between the mutual inclinations of systems with and without flybys.  Initial conditions for planet-planet scattering studies with very small initial inclinations may not be realistic. 

\item Both high and low eccentricities are affected by the presence of flybys, although the effects of flybys remains small compared with the initial  density of planets.  The median eccentricities are 0.015, 0.038, 0.019, and 0.047 for 2P1F0, 3P1F0, 2P1F1, and 3P1F1, respectively.  

\item  \ACBc{Radial} mixing of planetary orbits takes place in all simulations.  Wide-orbit planets can be placed on moderate orbits, and moderate-period planets can be placed on very long-period orbits.  Observing a planet at a given location in a disk is not by itself indicative of where and/or how it formed. Moreover, the scattering history of a planet can be complex, with the possibility that some planets will spend time in both short and long-period orbits during the system's evolution.

\item The scattering process can lead to very extreme outcomes, including turning a wide-orbit planet into a hot Jupiter.  In the four-planet simulations without flybys, nearly 3\% of the systems have a planet that at some point has a pericenter inside 0.1 AU.  In all cases the planets are initially at distances greater than 10 AU, and several are planets that are initially at distances of 100 AU. The planets that are scattered to such small pericenters are preferentially the lower mass planets in the simulations. We run a subset of the flyby simulations using a tidal damping model (2P1F1TD and 3P1F1TD), and show an example of a hot Jupiter that is formed from the inward scattering of a planet that is initially at 100 AU. The planet is also highly inclined, with an end inclination relative to the $x-y$ plane $i\sim 70^{\circ}$, \ACBc{a semi-major axis $a=0.019$, and an $e=0.013$}.
% If wide-orbit planets are typically born with high specific entropy, then their radii may be inflated at the time of tidal capture relative to planets that are born on lower entropy states, possibly reflecting a disk instability and core-accretion divide.  

\item The inclination distribution relative to the $x-y$ plane, here assumed to be normal to the stellar spin axis, is large for the planets that penetrate 0.1 AU at the time of smallest pericenter.  At least half of these planets have inclinations greater than $40^{\circ}$.

\item Stellar flybys can directly cause ejections of planets for $\qf$ that are within $\sim 2.5$ times the outermost planet's semi-major axis, which is consistent with results in the literature.  The frequency of ejections is strongly dependent on the initial density of planets in the system.  After weighting the effects of flybys to account for the expected encounter rates and including a distribution of planetary system sizes, we find that $\sim 30$ to 40\% of systems experience at least one ejection, with the typical ejection outcome leading to two-planet systems, consistent with previous work.   A few to about 8\% of systems have ejections that are induced by flybys, demonstrating that auto-ionization is the dominant mechanism for forming free-floating planets.  In 2P1F0, the fraction of planets ejected per system is 0.08, while this number jumps to $\sim 1$ for 3P1F0.  Flybys do increase the total number of ejected planets, with a perturber's influence being greatest on 2P1F1.   The total number of ejected planets per system is 0.21 and 1.2 for 2P1F1 and 3P1F1, respectively.

\item The dynamical signatures for long-period planets that are born {\it in situ} versus those that are scattered onto long periods are distinct due to differences in the expected eccentricity distributions.  Limited observations of relative proper motions between a companion and the host star may be able to constrain the contribution of {\it in situ} formation of planets on wide orbits, even using one component of the proper motion.  

\item Planet-planet scattering in systems where one planet was originally on a wide orbit can give rise to planets on very long-period orbits ($a\sim 1000$ AU) with pericenters $\sim 100$ AU.  In the simulations explored here, the fraction of systems that have a planet with a radial separation $>90$ AU at the end of the simulation are 0.97, 0.92, 0.85, and 0.83 for 2P1F0, 2P1F1, 3P1F0, and 3P1F1, respectively. If planets that were initially on a wide orbit are excluded, the frequency of planets with radial separations $>90$ AU ranges from 1 to 20 \% from 2P1F0 to 3P1F1.  

\end{enumerate}

%-----------------------------------------------------------------------------
\section{Acknowledgement}
We thank Dimitri Veras for sharing his scattering results.  We thank Fred Rasio and Smadar Naoz for helpful discussions\ACBc{, and we thank Fred Adams and 
the anonymous referee for comments that improved this manuscript.}
A.C.B.Õs support was provided by a contract with the California
Institute of Technology (Caltech) funded by NASA through
the Sagan Fellowship Program.
M.J.P.'s and E.B.F.'s contributions are supported by the National
Science Foundation under grant no.~0707203 and is also based
on work supported by NASA Origins of Solar Systems grant
NNX09AB35G.  T
The authors acknowledge the University of Florida High-Performance Computing Center for providing computational resources that contributed to the research results reported here.

%-----------------------------------------------------------------------------
% Appendix
\appendix
%-----------------------------------------------------------------------------

%-----------------------------------------------------------------------------
\section{Energy Conservation}\label{APP:ENERGY}

\begin{figure}
\begin{center}
\includegraphics[width=8.5cm]{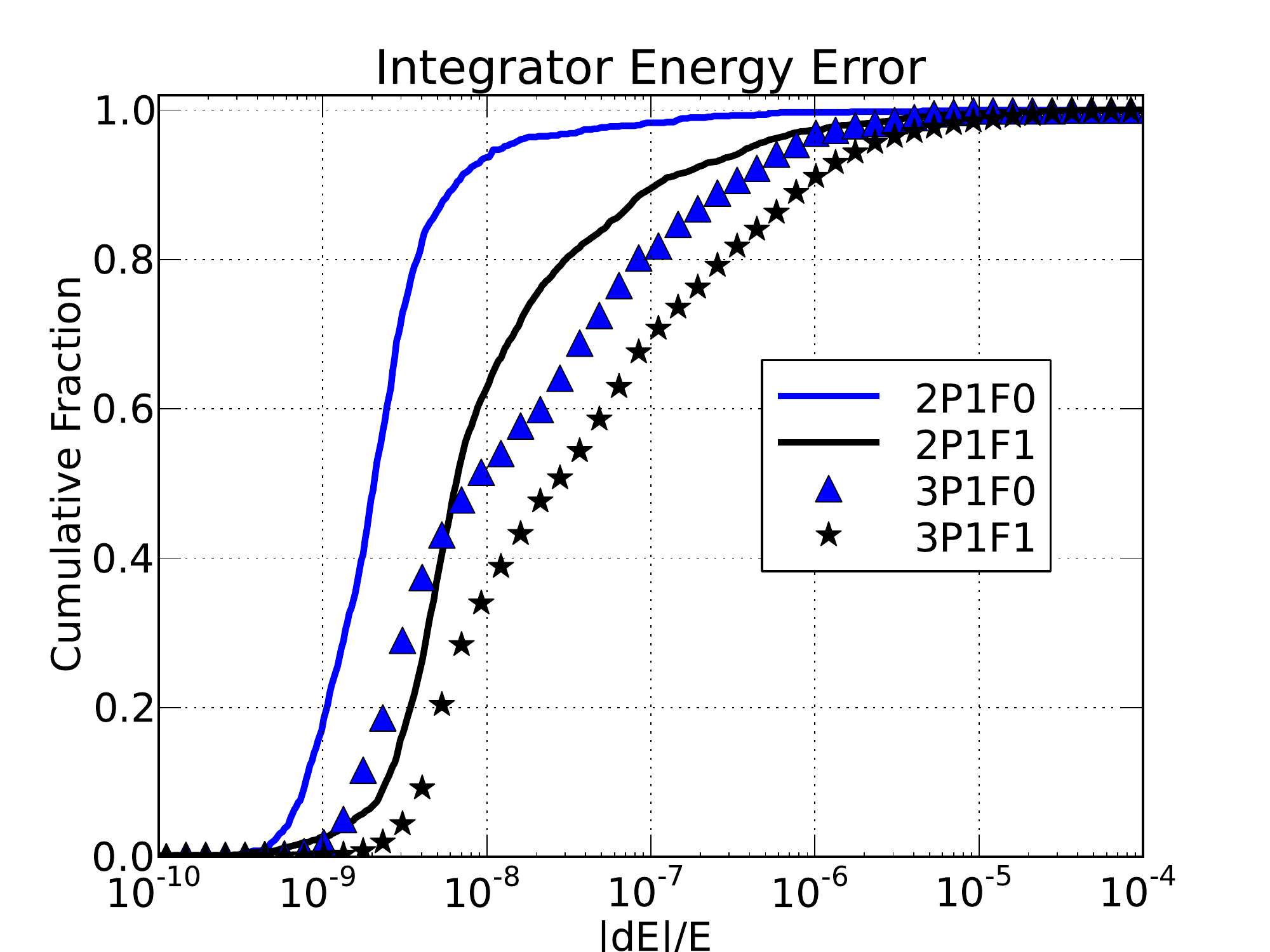}
\caption{Cumulative histograms of the integrator energy conservation for the simulations.  The median energy error for all simulations is smaller than  $dE/E \sim 10^{-7}$, and almost all systems have energy conservation smaller than $10^{-5}$.  There are seven systems in the simulations with flybys (2P1F1 and 3P1F1 combined) that have energy conservation worse than $10^{-4}$, but they do not change the general results. To make sure that systems with very poor energy conservation are not biased toward systems of interest, we list the median energy errors for all systems that have a planet with $q<0.1$ AU at some time during the system's evolution, which are $2\times10^{-7}$, $1\times10^{-6}$, $7\times10^{-7}$, and $1\times10^{-6}$ for 2P1F0, 2P1F1, 3P1F0, 3P1F1, respectively. }
\label{fig:app_energy}
\end{center}
\end{figure}

\ACBc{In Figure \ref{fig:app_energy}, we show cumulative histograms of the integrator energy conservation for the simulations.  The median energy error for all simulations is  $<\vert dE\vert /E \sim 10^{-7}$, and almost all systems have energy conservation $<10^{-5}$.  There are seven systems in the simulations with flybys (2P1F1 and 3P1F1 combined) that have energy conservation  $>10^{-4}$, but they do not change the general results. To make sure that systems with very poor energy conservation are not biased toward systems of interest, we list the median energy errors for all systems that have a planet with $q<0.1$ AU at some time during the system's evolution, which are $2\times10^{-7}$, $1\times10^{-6}$, $7\times10^{-7}$, and $1\times10^{-6}$ for 2P1F0, 2P1F1, 3P1F0, 3P1F1, respectively. We focus on these simulations because they all have experienced strong scattering events.  The maximum tolerable error can be estimated from the most extreme mass ratio in the problem, which implies that we require error conservation to be $<<10^{-3}$.  The median error for systems with some of the strongest scattering events meets this criterion. }

%-----------------------------------------------------------------------------
% Figures
%-----------------------------------------------------------------------------

%%1

%%2

%3

%4

%5

%6

%7

%8

%9

%10

%11

%12

%13

%14
%15

% ------------------------------------------------------------
% Figure for appendix is at the very bottom
% ------------------------------------------------------------

% ------------------------------------------------------------
% Tables
% ------------------------------------------------------------

\clearpage
\begin{table}
\begin{center}
\caption{ Definitions of symbols and abbreviations.\label{tab:definitions}}
\scriptsize
\begin{tabular}{l p{10cm} l p{1.5cm}}\hline\hline

Symbol 			& Definition 												& Units 					& Definition Section 		\\
\tableline
$a, a_i$ 			& Planetary Semi-major Axes 									& AU						& \ref{sec:numerics} 							\\
$e$ 				& Planetary Eccentricity										& AU						& \ref{sec:numerics}  					 		\\
$i$ 				& Planetary Inclination										& AU						& \ref{sec:numerics} 							\\
$q,\,q_{flyby}$		& Pericenter, Pericenter of Stellar Flyby							& AU						& \ref{sec:encounter_freq}					\\
$N$				& Number of Stars in Cluster									& - 						& \ref{sec:encounter_freq}					\\
$N_0 \& N_1$		& Min \& Max Values of N considered							& - 						&  \ref{sec:encounter_freq}					\\
$Q$				& Viral Parameter: $=\frac{\textrm{Total Kinetic Energy}}{\textrm{Total Potential Energy}}$	& -		&  \ref{sec:encounter_freq}					\\
$r_c$			& Core radius of stellar cluster									& pc						&  \ref{sec:encounter_freq}					\\
$\Gamma(q,N)$	& Rate of encounters with pericenters $<\qf$ in a cluster of size N			& \# star$^{-1}$ Myr$^{-1}$ 	&  \ref{sec:encounter_freq}					\\
$\gamma(N)$		& Encounter rate exponent									& - 						&  \ref{sec:encounter_freq}					\\
$\xi_m$			& Cluster stellar-mass function									& -						&  \ref{sec:encounter_freq}					\\
$\xi_N$			& Cluster stellar-number function								& -						&  \ref{sec:encounter_freq}					\\
$\eta$			& Average \# of encounters per star in $\Delta t$					&						&  \ref{sec:encounter_freq}					\\
$\eta^\prime$		& Fraction of field stars experiencing at least 1 encounter in $\Delta t$		&						&  \ref{sec:encounter_freq}					\\
$\Delta t$			& Typical interaction period within a cluster.  Typically 10.						 &		Myr				& \ref{sec:encounter_freq}					\\
$\frac{\eta}{\eta^\prime}$	& \# of encounters for systems with $>1$ encounter				&						&  \ref{sec:encounter_freq}					\\
$\Gamma^\prime$	& Rate of Stellar Flybys										&						& \ref{sec:sensitivities}					\\
$n$				& Cluster stellar density										&						&  \ref{sec:sensitivities}					\\
$\sigma$			& Interaction cross section									&						&  \ref{sec:sensitivities}					\\
$v$				& Average Speed of Star in Cluster								& 						&  \ref{sec:sensitivities}					\\
$m$				& Characteristic mass of Star in Cluster							& 						&  \ref{sec:sensitivities}					\\
$R_{sys}$			& Outer Radius of Planetary System/Initial Disk Size					& AU						&  \ref{sec:sensitivities}					\\
$r_0$			& Typical cluster size scale 									& pc						&  \ref{sec:sensitivities}					\\
$R_H$			& Mutual Hill Radius											& AU						& \ref{sec:scattering_design}					\\
$M_{\odot}$		& Solar Mass												& 						& \ref{sec:scattering_design} 					\\
$K$				& Planetary Separation in units of Mutual Hill Radii					&						& \ref{sec:scattering_design}					\\
$i_{median}$		& Median mutual planetary inclination 							&						& \ref{sec:excitation_discussion_100}					\\
$e_{median}$		& Median mutual planetary eccentricity 							&						& \ref{sec:excitation_discussion_100}					\\
$v_K$			& Keplerian Orbital Velocity									&						& \ref{sec:excitation_discussion_100}					\\
$v_r$			& Radial Velocity											&						& \ref{sec:excitation_discussion_100}					\\
$v_{\rm phi}$		& Azimuthal Velocity									&						& \ref{sec:excitation_discussion_100}					\\
$v_c$			& Circular Speed											&						& \ref{sec:excitation_discussion_100}					\\
$\delta v_r$		& Radial Velocity Dispersion									&						& \ref{sec:excitation_discussion_100}					\\
$\delta v_z$		& Azimuthal Velocity Dispersion								&						& \ref{sec:excitation_discussion_100}					\\
$q_{large}$		& Flyby $q$ for which all stars are expected to experience one encounter&						& \ref{sec:excitation_discussion_100}					\\
$f_{survive}$		& Fraction of systems that retain all their planets 					&						& \ref{sec:excitation_discussion_100}					\\
$J$				& Total Angular Momentum of a Cloud Core						&						& \ref{sec:excitation_discussion_general}					\\
$\mathcal{G}$		& Velocity Gradient of Cloud	 Core							& $km\,s^{-1}\,pc^{-1}$		& \ref{sec:excitation_discussion_general}						\\					\\
$M_{vir}$			& Viral Cloud Core Mass										&						& \ref{sec:excitation_discussion_general}						\\
$N_{pl}(a)$		& Planetary Semi-major Axis Distribution							&						& \ref{sec:constrain_wideorbits}				\\
$\mu$			& $G(m_{\rm star}+m_{\rm planet})$								&						& \ref{sec:constrain_wideorbits}			\\
$$				&														&						&					\\
\hline
$2P1F0$			& Simulations with 2 planets interior to 1 wide-orbit planet, with NO flyby		& -					& \ref{NB:ICs}					\\
$2P1F1$			& Simulations with 2 planets interior to 1 wide-orbit planet, with a stellar flyby	& -					& \ref{NB:ICs}					\\
$3P1F0$			& Simulations with 3 planets interior to 1 wide-orbit planet, with NO flyby		& -					& \ref{NB:ICs}					\\
$3P1F1$			& Simulations with 3 planets interior to 1 wide-orbit planet, with a stellar flyby	& -					& \ref{NB:ICs}					\\
$3P1F1C$		& Simulations with 3 planets interior to 1 wide-orbit planet, with a stellar flyby $\qf$ distribution of $\gamma=1.3$ & -					& \ref{NB:ICs}				\\
$2P1F1TD$			& Simulations with 2 planets interior to 1 wide-orbit planet, with a stellar flyby	and tidal damping & -					& \ref{sec:hotjupiters}\\
$3P1F1TD$			& Simulations with 3 planets interior to 1 wide-orbit planet, with a stellar flyby	and tidal damping & -					& \ref{sec:hotjupiters}\\
\hline
\end{tabular}
% Any table notes must follow the \end{tabular} command.
%\tablenotetext{a}{Sample footnote for table~\ref{tbl-2} that was generated with the \LaTeX\ table environment}
%\tablecomments{We might want to change the final column so that it refers to the section, given that the page number is going to change multiple times as we go to print versions, etc.}
\end{center}
\end{table}

\begin{table}
\begin{center}
\caption{ The fraction of field stars that have had a close encounter $<\qf$  for different $\qf$ and range of cluster sizes $N_0$ to $N_1$.  The column $\eta=\eta( \qf, N_0,N_1, \Delta t, \frac{d\xi_N}{dN})$ represents the average number of encounters per field star for flyby pericenter $\qf$.  The definition of $\eta$ (equation \ref{eq:eta}) includes stars that have had multiple encounters.  The quantity $\eta'$ does not include multiple encounters, so it represents the fraction of field stars that have had at least one encounter.  The average number of encounters among field stars that have had at least one encounter is given by the ratio of $\eta$ to $\eta'$. We use $\Delta t=10$ Myr for these calculations.  See Section \label{sec:encounter_freq} for more details. 
 \label{table:encounter_probabilities}}
\begin{tabular}{ l l l l l l }
$\qf ($AU) & $N_0$ & $N_1$ & $\eta$ & $\eta'$ & $\eta/\eta'$ \\\hline\hline
100 & 10 & $10^4$ & 0.18 & 0.18 & 1\\
100 & 30 & $10^4$ & 0.098 & 0.098 & 1\\
100 & 100 & $10^4$ &  0.044 & 0.044 & 1 \\\hline

200 & 10 & $10^4$ & 0.38 & 0.34 & 1.1 \\
200 & 30 & $10^4$ & 0.22 & 0.22 & 1\\
200 & 100 & $10^4$ & 0.011  & 0.11 & 1\\\hline

300 & 10 & $10^4$ & 0.59 & 0.44& 1.3 \\
300 & 30 & $10^4$ & 0.35 & 0.34 & $\sim1$ \\
300 & 100 & $10^4$  & 0.19 & 0.19 & 1\\\hline

1000 & 10 & $10^4$ &2.3 & 0.82 & 2.8 \\
1000 & 30 & $10^4$ &1.5 & 0.79 & 1.9 \\
1000 & 100 & $10^4$ &1.0 & 0.74 & 1.4 \\\hline\hline

\hline
\end{tabular}
\end{center}
\end{table}

\begin{table}
\begin{center}
\caption{The expectation values for the maximum mutual inclination and eccentricity of planets in a given system for two different assumptions for the system size $R_{\rm sys}$.  The cluster size limits are taken to be between $N_0$ and $N_1$ for the integration of equation (\ref{eqn:expected_i}). \label{table:expectation_values}}
\begin{tabular}{ l l l l l l }
 $N_0$ & $N_1$ & $\left< i\right>$ (deg,rad) & $\left< e \right>$ & $\left< i\right>$ (deg,rad) & $\left< e \right>$ \\\hline\hline
 & & \multicolumn{2}{c}{100 AU} & \multicolumn{2}{c}{30 AU} \\\hline
 10 & $10^4$ & 5.4,~0.094 & 0.083 & 1.8,~0.031 & 0.043 \\
 30 & $10^4$ & 3.5,~0.061 & 0.056 & 1.0,~0.017 & 0.035\\
 100 & $10^4$ &  1.9,~0.033 & 0.042 & 0.49,~0.0086  & 0.029\\\hline

\hline
\end{tabular}
\end{center}
\end{table}

\clearpage

\begin{sidewaystable}
\begin{center}
\footnotesize
\caption{Median inclinations of all bound planets for the initial and final states in the simulations.  Planet-planet scattering alone leads to a two-peak distribution in inclinations, with a low-inclination peak that reflects the initial state of the system and a broad, high-inclination peak due to scattering (see Fig.~4-5).  The two peaks are separated at  $i\sim0.3$ degrees for the relative to the $x-y$ plane and the mutual inclination distributions. The presence of perturbing stars smears out this two-peak distribution. The initial median inclinations are less than $0.05^{\circ}$ because all planets at 100 AU are set with zero initial inclination. In the last two columns, the median eccentricities are also given.}
\begin{tabular}{l cccccccccc }\\\hline
& \multicolumn{2}{c}{IC (deg)} & \multicolumn{2}{c}{Final (deg)} &  \multicolumn{2}{c}{Final (deg)} &  \multicolumn{2}{c}{Final (deg)} & &\\
Name & $x-y$ Plane & Mutual & $x-y$ Plane & Mutual &  $x-y$ Plane & Mutual  & $x-y$ Plane & Mutual  &  \multicolumn{2}{c}{Eccentricity}\\
& & & & & \multicolumn{2}{c}{$i>0.3^{\circ}$}  &  \multicolumn{2}{c}{$\qf>300$ AU} & All &  {$\qf>300$ AU} \\\hline 
2P1F0 & 0.024 & 0.073 & 0.038 & 0.080 & 2.4 & 3.5 & - & - & 0.015 & - \\
3P1F0 & 0.034 & 0.085 & 0.075 & 0.12 & 5.9 & 11 & - & - & 0.038 & -\\
2P1F1 & 0.025 & 0.074 & 0.18 & 0.24 &  - & - & 0.13 & 0.19 & 0.019 & 0.017 \\
3P1F1 & 0.034 & 0.087 & 0.65 & 0.86 & - & - & 0.33 & 0.45 & 0.047 & 0.040\\
3P1F1C & 0.033  & 0.086 & 0.67 & 0.96 & - & - & - & - & 0.049 & - \\
\end{tabular}
\label{table:median_inclinations}
\end{center}
\end{sidewaystable}

\begin{table}[h]
\begin{center}
\caption{Histogram data for the distribution $R_J$ (left) and $J$ (right) in low-mass cloud cores, based on the \citet{caselli_etal_2002_apj_572} observations.  The data are for the 20 sources (their Table 5) that have enough information to estimate $R_J$, the angular momentum barrier for a rotating, collapsing cloud. Note that bin values for the $J$ distribution will not correspond to the $R_J$ bin because conversion from $J$ to $R_J$ requires the mass of the cloud.  Here $F_{\rm 40 AU bin}$ and $F_{0.4dex}$ are the fraction of sources that fall within the given bin. The final row in the table shows the average of all the bins weighted by $F_{\rm 40 AU bin}$.  No weighing for the angular momentum distribution is given. \label{table:system_sizes}}
\begin{tabular}{| l l l l l l || l l |}\hline
$R_J$ (AU) & $F_{\rm 40~AU~bin}$  (AU) & $N_0$ & $\left<i_{\rm median}\right>$ &$\left<e_{\rm meidan}\right>$ & $f_{\rm survive}$ & $\log_{10}(J(\rm cm^2s^{-2}))$ & $F_{\rm 0.4 dex}$\\ \hline\hline
& & 10  & 1.2 & 0.037 & 0.67 &  &  \\ 
20 &0.6 & 30 & 0.68 & 0.031 & 0.68 & 19.9 & 0.05 \\ 
 &  & 100 & 0.33 & 0.028 & 0.69 & & \\ \hline
& & 10 & 3.5 & 0.061 & 0.61 &  & \\
60  &0.1 & 30 & 2.1 & 0.045 & 0.65 &  20.3 & 0.2 \\
& & 100 & 1.0 & 0.034 & 0.68 &  & \\\hline
& & 10 & 5.4 & 0.083 & 0.55 & & \\
100& 0.2 & 30 & 3.5 & 0.060 & 0.61 & 20.7 & 0.35 \\
 & & 100 & 1.9 & 0.042 & 0.66 & & \\\hline
& & 10 & 7.1 & 0.010 & 0.50 & & \\
140& 0.05 & 30 & 4.8 & 0.075 & 0.57 & 21.1 & 0.35 \\
& & 100 & 2.8 & 0.051 & 0.63 & & \\\hline
& & 10 & 8.5 & 0.012 & 0.47 & & \\
180& 0.05 & 30 & 6.1 & 0.089 & 0.53 & 21.5 & 0.05 \\
& & 100 & 3.7 & 0.061 & 0.60 & & \\\hline
& & 10 & 2.9 & 0.056 & 0.62 & & \\
\multicolumn{2}{|c}{Weighted Average} & 30 & 1.9 & 0.15 & 0.65 & & \\
& & 100 & 1.0 & 0.034 & 0.68 & & \\\hline

\end{tabular}
\end{center}
\end{table}

% ------------------------------------------------------------
% Figure for appendix
% ------------------------------------------------------------

%\bibliographystyle{apj}
%\bibliography{master_ref}

\end{document}